\documentclass[aps,prb,twocolumn,superscriptaddress,floatfix]{revtex4-2}
\usepackage{graphicx,graphics}
\usepackage{dcolumn}
\usepackage{amsmath,amssymb,amsfonts,nccmath}
\usepackage{latexsym,verbatim}
\usepackage{bm}
\usepackage{color}
\usepackage{ulem}
\usepackage{soul}
\usepackage{braket}
\usepackage[percent]{overpic}
\usepackage[breaklinks=true,colorlinks,citecolor=blue,linkcolor=blue,urlcolor=blue]{hyperref}
\usepackage{lipsum}
\usepackage{enumitem}
\usepackage[T1]{fontenc}

\usepackage{xprintlen}

\makeatletter
\renewcommand\@make@capt@title[2]{%
 \@ifx@empty\float@link{\@firstofone}{\expandafter\href\expandafter{\float@link}}%
  {\textbf{#1}}\@caption@fignum@sep#2\quad
}
\makeatother

\DeclareMathOperator{\sign}{sgn}

\begin{document}
\title{Tunable interface states between Floquet-Weyl semimetals}
%

%
\author{F. Bonasera}
\affiliation{Dipartimento di Fisica e Astronomia ``Ettore Majorana'', Universit\`a di Catania, Via S. Sofia 64, I-95123 Catania,~Italy}
\affiliation{INFN, Sez.~Catania, I-95123 Catania,~Italy}
\author{S.-B. Zhang}
\affiliation{Institute for Theoretical Physics, University of Zurich, 8057 Zurich, Switzerland}
\author{L. Privitera}
\affiliation{Institute for Theoretical Physics and Astrophysics, University of Würzburg, D-97074 Würzburg, Germany}
\affiliation{Corporate Research, ABB Switzerland Ltd., Segelhofstrasse 1K, Baden-Dättwil 5405, Switzerland}
\author{F.M.D. Pellegrino}
\affiliation{Dipartimento di Fisica e Astronomia ``Ettore Majorana'', Universit\`a di Catania, Via S. Sofia 64, I-95123 Catania,~Italy}
\affiliation{INFN, Sez.~Catania, I-95123 Catania,~Italy}
\affiliation{CNR-IMM, Via S. Sofia 64, I-95123 Catania, Italy}
\begin{abstract}
Weyl semimetals and nodal line semimetals are characterized by linear electronic bands touching at zero-dimensional points and one-dimensional lines, respectively. Recently, it has been predicted that nodal line semimetals can be driven into tunable Floquet-Weyl semimetals by circularly polarized light.
Here, we study the occurrence of interface states between two regions of a nodal line semimetal illuminated by two beams of light with opposite circular polarizations. 
Within a minimal model, we find remarkable modifications of the energy structure by tuning the polarized light, such as the possible generation of Van Hove singularities.
Moreover, by adding a $\delta$-doping of magnetic impurities along the interfacial plane, we show the occurrence of a switchable and topologically non-trivial, vortex-like pseudo-spin pattern of the interface states.
\end{abstract}
\maketitle

\section{Introduction}
\label{sec:intro}

The field of research in topological materials has seen an outburst since the discovery of the first quantum state to break the Landau paradigm on the classification of phases of matter~\cite{Klitzing_1980}. A new characterization of these materials through topology has been established\cite{Laughlin_1981,Thouless_1982,Kane_2005_b,Thouless_1998,Hasan_2010,Qi_2011,
Armitage_2018,Fang_2016,Chiu_2016}.
One of the most dramatic effects of a non-trivial topology in many materials is the existence of metallic, dissipationless and protected states that occur at the boundaries of the sample. This property is known as bulk-boundary correspondence, it has firstly been discovered in the context of the integer quantum Hall effect\cite{Hatsugai_1993}, and it was then predicted and observed in topological insulators~\cite{Kane_2005_a,Bernevig_2006,Konig_2007,Roth_2009,Hasan_2010,Qi_2011,Chiu_2016}
and in other gapless materials, such as the Dirac semimetals~\cite{Murakami_2007,Young_2012,Liu_2014,Neupane_2014,Borisenko_2014,Armitage_2018,
Chiu_2016},
the nodal line semimetals~\cite{Burkov_2011_b,Xie_2015,Fang_2015,Kim_2015,Bian_2016_a,Bian_2016_b,Chan_2016_a,
Yu_2015,Chiu_2016,Yu_2017,Fang_2016}
(NLSMs) and the Weyl semimetals~\cite{Nielsen_1983,Burkov_2011_a,Wan_2011,Lv_2015,Huang_2015,Xu_2015,Rao_2016,Yan_2017_a,
Armitage_2018,Chiu_2016,Pellegrino_2015_a,Andolina_2018_a}
(WSMs).
Specifically, the NLSM
is characterized by linear valence and conduction bands that cross each other along a one-dimensional curve in the three-dimensional Brillouin zone at the Fermi energy
\cite{Chiu_2016,Fang_2016},
whereas in the WSM the linear valence and conduction bands touch each other in single points, known as Weyl points (WPs). 
Each WP is either a source or a sink of a quantum of Berry flux, and a couple of WPs produces characteristic zero-energy lines in the surface energy spectrum, known as Fermi arcs~\cite{Vanderbilt_2018,Armitage_2018,Rao_2016,Yan_2017_a,Chiu_2016}.

Recently, the interest in topological materials has moved towards their possible applications for quantum technologies~\cite{Kitaev_1997,Kitaev_2001,Nayak_2008,Moore_2010,Beenakker_2013,Sarma_2015,Sato_2017,Lian_2018,Chen_2020}.
The robustness of the topological properties against weak perturbations makes this materials good candidates as efficient platforms for quantum computing~\cite{Sato_2017}.
In this framework, the problem of controlling and engineering these topological phases of matter rises. In particular, the Floquet engineering of a system through a periodic drive has been the subject of a wide theoretical investigation~\cite{Lindner_2011,Kitagawa_2010,Oka_2009,Inoue_2010,Gu_2011,Kitagawa_2011,
Rudner_2013,Wang_2014,Hubener_2017,Chan_2016_b,Jiang_2011,Mahmood_2016,Lindner_2013,Gomez-Leon_2013,Delplace_2013,Oka_2019_a,Kitagawa_2010,Rudner_2013,Carpentier_2015_a,
Harper_2020_a,Khemani_2016_a,
Else_2016_a,Roy_2017_a,Potter_2016_a,Else_2016_b,vonKeyserlingk_2016_a,
Fidkowski_2019_a,Roy_2017_b,Wang_2014,Hubener_2017,Delplace_2013,Kim_2015,
Zhang_2016_b,Chen_2018,
Yan_2017_b,Chan_2016_c,Yan_2016,Taguchi_2016,Narayan_2016,Okugawa_2017}, and
experimental studies have also been performed
\cite{Wang_2013,Zhang_2014_a,Mahmood_2016,Rudner_2020_a,Eckardt_2017_a,Wintersperger_2020,delaTorre_2021_a}.
The concept of Floquet engineering has also been recently applied to topological systems with the aim of, for example, creating anomalous topological states with no static counter part
\cite{Kitagawa_2010,Rudner_2013,Carpentier_2015_a,Harper_2020_a,Khemani_2016_a,
Else_2016_a,Roy_2017_a,Potter_2016_a,Else_2016_b,vonKeyserlingk_2016_a,
Fidkowski_2019_a,Roy_2017_b,Dag_2022}
and, in particular, of inducing topological phase transitions
\cite{Cayssol_2013,Kundu_2014,Wang_2014,Hubener_2017,Delplace_2013,Kim_2015,Zhang_2016_b,Chen_2018,Yan_2017_b,
Chan_2016_c}.
Specifically, it was predicted that irradiating a NLSM with circularly polarized light induces, to first order in the inverse frequency of light $1/\omega$, a transition to a WSM phase with tunable WPs
\cite{Yan_2016,Taguchi_2016,Narayan_2016,Okugawa_2017}.
Recently, 
interface electron systems with topological materials have attracted attention~\cite{Ishida_2018_a,Dwivedi_2018}, also those generated by a dynamic drive~\cite{Calvo_2015_a,Islam_2019_a}.

In this work, we study the emergence of electronic surface states at the interface between two half-spaces of a NLSM irradiated by two monochromatic light beams with opposite circular polarizations, respectively.
Here, we show how two different light intensities can modify the electronic band structure up to introducing a Van Hove singularity (VHS) in the density of states of the interface system. A VHS is a logarithmic divergence in the density of states generally caused by a saddle point in the energy spectrum of 2D systems~\cite{VanHove_1953,Grosso_2000}. When VHSs lie at the Fermi energy, they enhance electron interactions and can cause electronic instabilities: they can induce phenomena such as superconductivity~\cite{Kohn_1965,Hirsch_1986,Dzyaloshinskii_1987,Markiewicz_1991,Newns_1992,Markiewicz_1997,Gonzalez_2008,Li_2009,McChesney_2010,Kim_2018,Isobe_2018,Sherkunov_2018,Wu_2021,Wan_2022,Kang_2022}, charge density waves~\cite{Rice_1975,Cranney_2010,Sherkunov_2018} and spin density waves~\cite{Makogon_2011,Fradkin_2013,Isobe_2018,Liu_2018,Sherkunov_2018}. For these reasons, VHSs in topological systems have attracted great interest in the aim for exotic correlated quantum states~\cite{Xu_2015_b,Singh_2018,Ghosh_2019,Wang_2020,Wu_2021_b,Lee_2021,Hu_2022_a}, such as topological superconductivity~\cite{He_2014,Yao_2015,Hu_2022_b}. Moreover, we find that, by adding a narrow magnetic barrier at the interface between the two induced WSMs, it is possible to switch on/off a topological phase of the interface system by using suitable different light intensities.

This paper is organized as follows. In Sec.~\ref{sec:model} we introduce a simple two-band model used to describe a $z$-symmetric NLSM.
In Sec.~\ref{sec:driving}, following Ref.~\onlinecite{Yan_2016}, we show how, using the Floquet formalism in the high frequency limit~\cite{Eckardt_2015}, the circularly polarized light induces a transition from NLSM to WSM.
In Sec.~\ref{sec:interface}, we study the interface states which lay along the boundary between two induced WSMs. Within the two-band model, the interface eigenstates have a spinorial form, where the components represent the orbital degree of freedom. Here, we focus on energy dispersion and pseudo-spin texture $\langle \bm{\sigma}\rangle$ evaluated on the interface states, where the Pauli matrix vector $\bm{\sigma}$ acts on the orbital subspace.
Finally, we show in Sec.~\ref{sec:interface_magnetic} the topological effects on the interface system of introducing a magnetic barrier between the two induced WSMs.

\section{Model}\label{sec:model}

\begin{figure}[t]
\centering
\begin{overpic}[width=0.98\columnwidth,trim={0 0cm 0 0cm}]{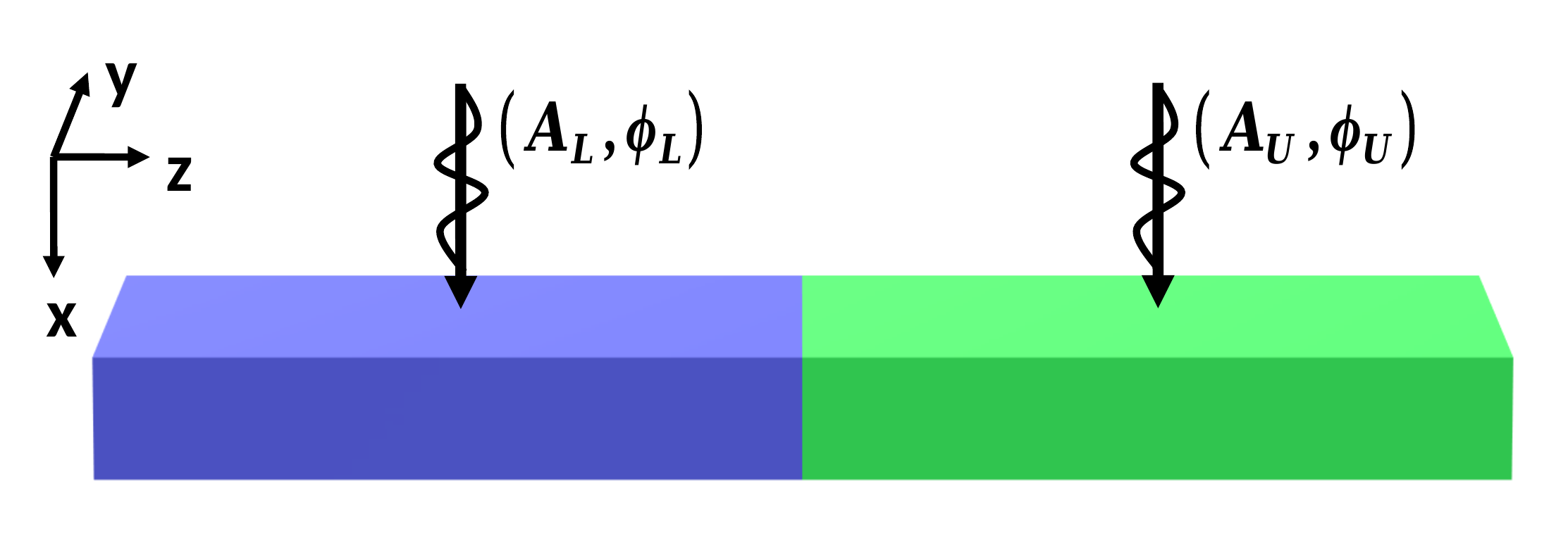}\end{overpic}
\caption{Schematic of the system. The lower, $z<0$, and upper, $z>0$, half-spaces of the infinite NLSM are irradiated by high frequency monochromatic lights of intensities  and polarizations $(A_{\rm L},\phi_{\rm L})$ and $(A_{\rm U},\phi_{\rm U})$, respectively.
}
\label{fig:setup}
\end{figure}
In this work, we study the interface states which emerge at the boundary plane of two half-spaces of an infinite NLSM irradiated respectively by two light beams of opposite circular polarizations.
Figure~\ref{fig:setup} shows the setup analyzed, where the upper (lower) half-space $z<0$ ($z>0$) is represented by the blue (green) region of the NLSM, and it is irradiated by a light beam characterized by an intensity $A_{\rm L}$ ($A_{\rm U}$) and a phase $\phi_{\rm L}$ ($\phi_{\rm U}$).
Here, we focus on monochromatic light in the high frequency regime, such that we can use the high frequency expansion within the Floquet formalism~\cite{Eckardt_2015}. 
Without loss of generality, we consider a light drive polarized along the $y-z$ plane and described by the following vector potential
\begin{equation}\label{eq:VectorPotential}
\bm{A}(t)=A \left[ 0, \cos{\left(\omega t\right)},\sin{\left(\omega t +\phi\right)} \right],
\end{equation}
where $\phi=0$ ($\phi=\pi$) corresponds to the right (left) handed circular polarization.

To describe many topological materials at low energies, it is usually adopted the following model~\cite{Zhang_2009,Burkov_2011_b,Qi_2011,Kim_2015,Zhang_2016_a,Fang_2015} based on the Dirac Hamiltonian supplemented by quadratic corrections 
\begin{equation}\label{eq:H_Dirac_Full_Quadratic}
H(\bm{k}) = \left(m-\sum_{i=x,y,z} B_i k_i^2\right)\beta + \sum_{i=x,y,z} \alpha_i v_i k_i + \varepsilon\left(\bm{k}\right) \openone,
\end{equation}
where the Dirac matrices $\{\beta,\alpha_i\}$ square to one and anticommute with each other, $\openone$ is the identity matrix, $\bm{k}$ is the crystal momentum, $\varepsilon(\bm{k})$ is an electron-hole symmetry breaking term, and we have set $\hbar=c=k_B=1$. 
The values of the microscopic terms $m$, $\bm{B}$, $\bm{v}$ and $\varepsilon(\bm{k})$ depend on the specific topological material~\cite{Chan_2016_a,Xu_2017,Yu_2017}. 
The number of non zero components of $\bm{B}$ and $\bm{v}$ determines the dimension of the degeneracy subspace in the energy spectrum. Moreover, the non-trivial topology of systems described by the Hamiltonian \eqref{eq:H_Dirac_Full_Quadratic} is given by the band inversion that occurs from $|\bm{k}|=0$, $H(\bm{k})\to m \beta$, to $|\bm{k}| \to +\infty$, $H(\bm{k})\to-\sum_{i=x,y,z} B_i k_i^2 \beta$, with $mB_i>0$~\cite{Shen_2011}.
To analyze the interface states which lay along the plane $z=0$, without loss of generality, in the Hamiltonian expressed in Eq.~\eqref{eq:H_Dirac_Full_Quadratic} we disregard the quadratic momentum contribution along $z$-direction ($B_z=0$). 
This simplification allows using a fully analytical approach without missing any topological features of the electron system,  this is illustrated in appendix~\ref{sec:numerical}.
Therefore, the low-energy electronic properties of the NLSM are described by the following Hamiltonian
\begin{equation}\label{eq:H_NLSM}
    H(\bm{k}) = \left[m-B\left(k_x^2 + k_y^2\right)\right]\sigma_x + v k_z \sigma_z,
\end{equation}
where $\{\sigma_i\}$ are the Pauli matrices which act on an orbital subspace, and, for sake of simplicity, we have dropped the electron-hole symmetry breaking term $\varepsilon(\bm{k})$.
It is useful to rewrite the Hamiltonian \eqref{eq:H_NLSM} as
\begin{equation}\label{eq:H_NLSM_Dimensionless}
\mathcal{H}\left(\bm{p}\right) = \left[1-\left(p_x^2 + p_y^2\right)\right]\sigma_x + u p_z \sigma_z,
\end{equation}
where the energies are expressed in units of $m$, and the dimensionless momentum and velocity are defined as $\bm{p}=\bm{k}/b$ and $u=(b/m)v$, with $b=\sqrt{m/B}$.
The eigenenergies of the Hamiltonian~\eqref{eq:H_NLSM_Dimensionless} are expressed as
\begin{equation}\label{eq:H_NLSM_Eigenvalues}
    E_{\pm}(\bm{p}) = \pm \sqrt {\left[1-\left(p_x^2+p_y^2\right)\right]^2 +\left( u p_z\right)^2}~,
\end{equation}
where conduction and valence bands touch each other at the circular nodal ring defined by $p_x^2+p_y^2=1$, on the $p_z=0$ plane. Finally, we notice that the nodal ring is protected by the $z$-mirror symmetry, i.e. $(i\sigma_x)\mathcal{H}(\bm{p}_\perp,p_z)(-i\sigma_x) = H(\bm{p}_\perp,-p_z)$, where $\bm{p}_\perp=(p_x,p_y)$.

\section{Drive NLSM into FWSM}
\label{sec:driving}

Following Ref.~\onlinecite{Yan_2016}, within the Floquet formalism and accordingly to the high frequency expansion, we show here how a NLSM can be driven into a Floquet-Weyl semimetal (FWSM) by shining circularly polarized monochromatic light on it. 
In the presence of the vector potential of the form expressed in Eq.~\eqref{eq:VectorPotential}, by applying the Peierls substitution $\bm{p} \to \bm{p} + e\bm{A}(t)/b$, we obtain a time periodic Hamiltonian
\begin{align}\label{eq:H_NLSM_Periodic}
    \mathcal{H}(\bm{p},t) &= \left\{1-p_x^2
    -\left[p_y+e \Lambda \cos(\omega t)  \right]^2
   \right\}\sigma_x \\ &+ u \left[p_z +e \Lambda \sin{\left(\omega t +\phi\right)}\right]\sigma_z,\nonumber
\end{align}
where the dimensionless quantity $\Lambda = A/b$ is proportional to the light intensity $A$. 
The periodic Hamiltonian above can be expanded in Fourier series as $\mathcal{H}(\bm{p},t)=\sum_n \mathcal{H}_n(\bm{p})e^{-in\omega t}$, where
\begin{subequations}\label{eq:H_NLSM_Fourier}
\begin{align}
    \mathcal{H}_0  & =  \left[1-e^2\Lambda^2/2 - \left(p_x^2+p_y^2\right)\right]\sigma_x + u p_z \sigma_z  \\
    \mathcal{H}_{\pm 1} & = - e \Lambda \left(2p_y \sigma_x \pm i e^{\pm i\phi} u \sigma_z \right)/2 ,\\
    \mathcal{H}_{\pm 2} & = - e^2 \Lambda^2 \sigma_x /4~,
\end{align}
\end{subequations}
and $\mathcal{H}_{\pm n} = 0$ for $|n|>2$. 
Within the Floquet formalism, in the high frequency regime ($\omega \gg \left| m \right| $), we resort to a perturbative approach which describes the dynamics of the system by the time independent effective Hamiltonian~\cite{Eckardt_2015}
\begin{equation}\label{eq:H_Floquet_Effective}
    \mathcal{H}_{\rm{eff}}(\bm{p}) = \mathcal{H}_0 (\bm{p}) + \sum_{n\geq 1} \frac{\left[\mathcal{H}_{+n},\mathcal{H}_{-n}\right]}{n\Omega} + \mathcal{O}\left(\frac{1}{\Omega^2}\right),
\end{equation}
where $ \Omega = \omega/m $ is the frequency of the incident light in units of $m$.
Using the Fourier coefficients expressed in Eqs.~\eqref{eq:H_NLSM_Fourier}, we obtain the following commutators
\begin{subequations}
\begin{align}
	\left[\mathcal{H}_{+1},\mathcal{H}_{-1}\right] & = \frac{e^2 \Lambda^2}{4} \left[ 2p_y \sigma_x + i e^{i\phi} u \sigma_z, 2 p_y \sigma_x - ie^{-i\phi} \sigma_z \right] \nonumber \\
	& = \frac{e^2 \Lambda^2 u p_y}{2} \left( e^{i\phi} + e^{-i\phi} \right) i \left[ \sigma_z, \sigma_x \right] \nonumber \\
	& = - 2 e^2 \Lambda^2 u \cos\left(\phi\right) p_y \sigma_y, \\
	\left[\mathcal{H}_{+2},\mathcal{H}_{-2}\right] & = 0,
\end{align}
\end{subequations}
which lead to the effective Hamiltonian
\begin{equation}\label{eq:H_FWSM}
    \mathcal{H}_{\rm{eff}}(\bm{p}) = \left[\bar{p}^2 - \left( p_x^2 +p_y^2\right) \right] \sigma_x + \lambda p_y \sigma_y + u p_z \sigma_z,
\end{equation}
where
\begin{subequations}\label{eq:H_FWSM_Factors}
\begin{align}
    \bar{p} & = \sqrt{1- e^2 \Lambda^2 /2} \label{eq:H_FWSM_Factors_a} ,\\
    \lambda & = - 2 e^2 \Lambda^2 u \cos{\left(\phi\right)} /\Omega.\label{eq:H_FWSM_Factors_b}
\end{align}
\end{subequations}
The corresponding eigenenergies are
\begin{equation}\label{eq:H_FWSM_Eigenvalues}
   \mathcal{E}_{\pm}\left(\bm{p}\right) = \pm \sqrt{\left[\bar{p}^2-\left(p_x^2+p_y^2\right)\right]^2+\left(\lambda p_y\right)^2 + \left( u p_z \right)^2},
\end{equation}
which are degenerate only at two WPs placed at $\bm{P}_{\pm} = \left( \pm \bar{p},0,0\right)$. 
By linearizing the Hamiltonian expressed in Eq.~\eqref{eq:H_FWSM} around the WPs as $\mathcal{H}_{\rm{eff}}(\bm{q}_{\pm})=\sum_{ij} v_{ij} q_{i,\pm} \sigma_j$, with $q_{i,\pm}=p_i-P_{i,\pm}$, and by calculating the determinant of $v_{ij}$, we find the chiralities of the two WPs as $\chi_{\pm}= \sign[{\rm Det}(v_{ij})]=\pm\sign{\left(\cos\phi\right)}$. These are only determined by the polarization of the incident light~\cite{Yan_2016}. 
\section{Interface system}
\label{sec:interface}

In this section, we focus on the states which emerge at the interface between two regions of an infinite NLSM irradiated by two beams of light with opposite circular polarizations. 
In the high frequency regime,  we describe each region by an effective Floquet Hamiltonian, which is denoted as $H_{\rm U}$ ($H_{\rm L}$)  for the upper (lower) half-space $z>0$ ($z<0$).
The effective Floquet Hamiltonians are expressed as
\begin{equation}\label{eq:H_Setup}
	\begin{aligned}
	\mathcal{H}_j \left(\bm{p}\right)= \left[ \bar{p}_j^2 - \left(p_x^2 + p_y^2\right)\right]\sigma_x + \lambda_j p_y \sigma_y + u p_z \sigma_z
	\end{aligned}
\end{equation}
where
\begin{subequations}\label{eq:H_FWSM_Factors_LU}
\begin{align}
    \bar{p}_j & = \sqrt{1- e^2 \Lambda_j^2 /2} \label{eq:H_FWSM_Factors__LU_a} ,\\
    \lambda_j & = - 2 e^2 \Lambda_j^2 u \cos{\left(\phi_j\right)} /\Omega ,\label{eq:H_FWSM_Factors__LU_b}\\
    \Lambda_j&=\frac{A_j}{b},
\end{align}
\end{subequations}
$j\in\{\rm L,U\}$ denotes the half-space, and $\sign{\left(\lambda_{\rm U} \lambda_{\rm L}\right)} < 0$ because of the opposite circular polarizations of the two beams of light.
\begin{figure*}[t!] 
\centering
\begin{overpic}[width=0.99\columnwidth]{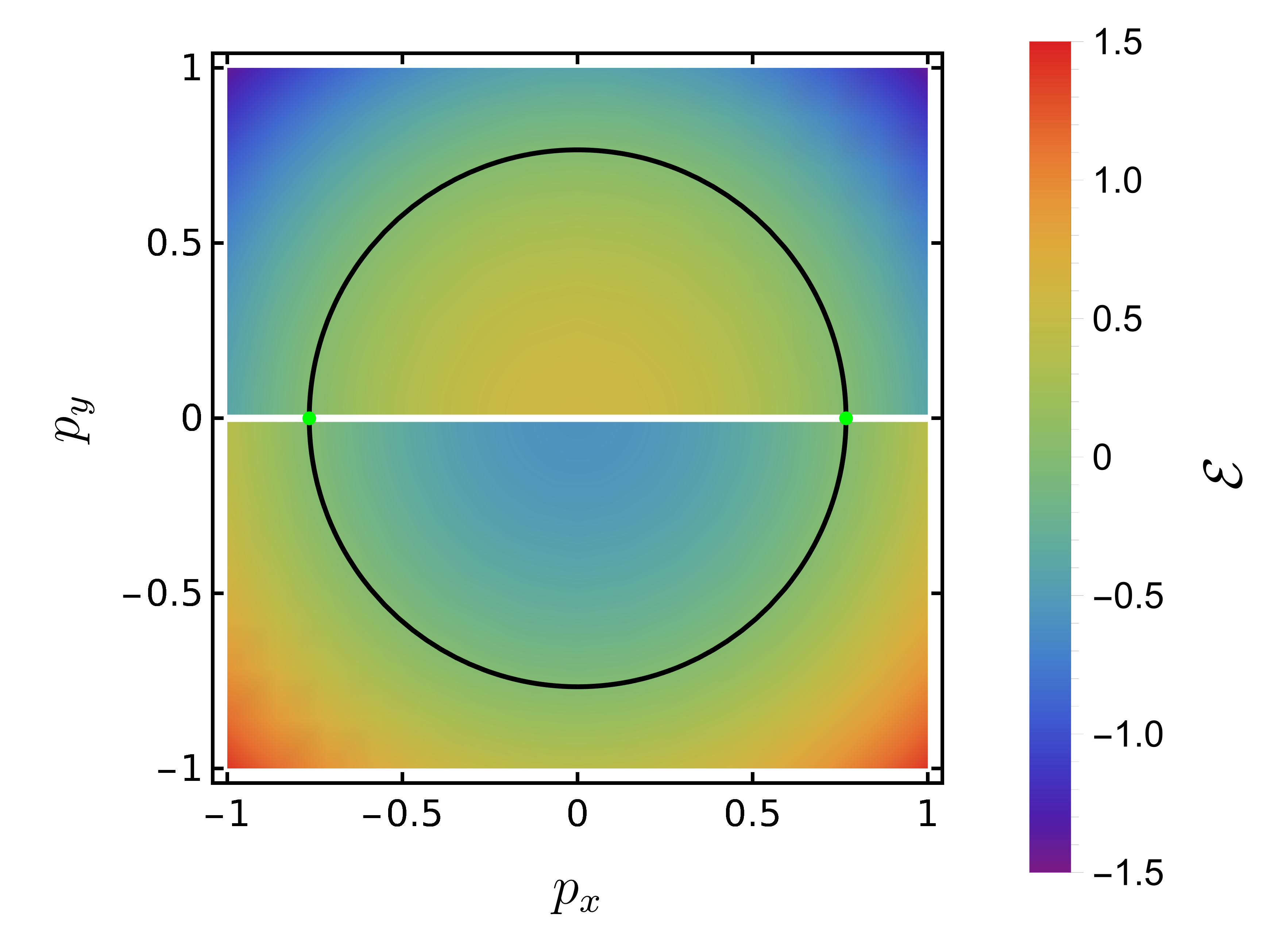}\put(4,75){(a)}\end{overpic}
\begin{overpic}[width=0.99\columnwidth]{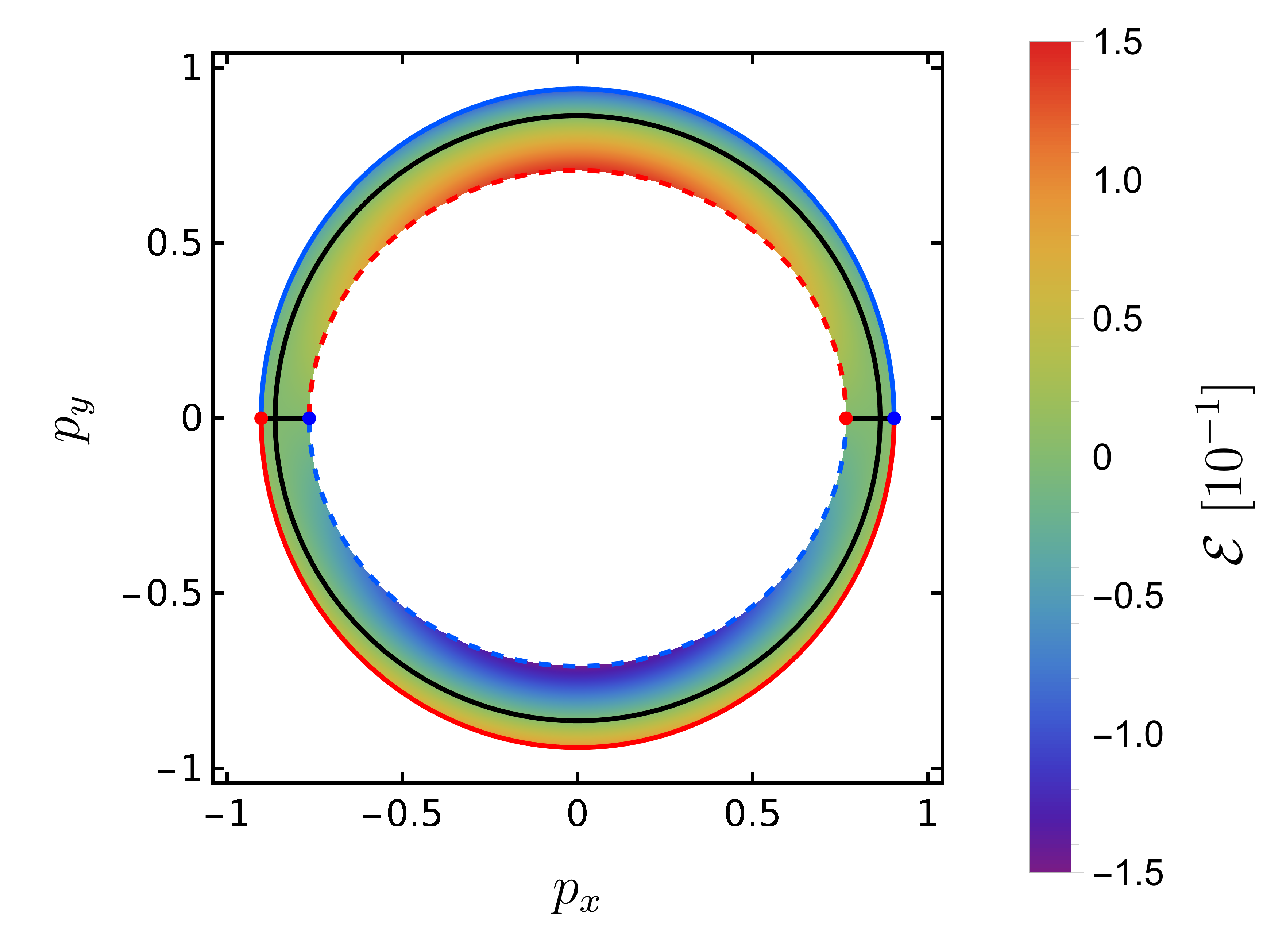}\put(4,75){(b)}\end{overpic}
  \caption{Density plots of the energy dispersion of the interface states as a function of the momentum $\bm{p}_\perp=(p_x,p_y)$. In both panels, the black lines represent the Fermi line ${\cal E}=0$, 
  and we set $u=1,\Omega=10$, $\phi_{\rm U} = \pi$ and $\phi_{\rm L}=0$.
  In panel~(a), the light intensities are identical, $\Lambda_{\rm U}=\Lambda_{\rm L}=3$. Here, each green dot represents two coincident WPPs with opposite chiralities, and the solid white line represents the locus where the interface electronic band merges into the bulk bands.
 In panel~(b), the light intensities are different,  $\Lambda_{\rm U}=2$ and $\Lambda_{\rm L}=3$.
Here, the boundary solid (dashed) lines are solutions of $\bar{\mu}_{\rm U}=0$  ($\bar{\mu}_{\rm L}=0$). In particular, the red (blue) lines describe the merging of the interface band into a conduction (valence) bulk band. Moreover, red and blue dots represent WPPs with positive and negative chirality, respectively.
  \label{fig:EnergyDispersionNonMagnetic}}
\end{figure*}
We note here that the assumption we make of an infinitely sharp interface between the two regions does not undermine the results of the study. Indeed, in this work we study topological interface modes which are a consequence of the topology mismatch between the two regions and are thus robust against the details of the system, such as those of the interface. For example, this was shown in Ref.~\cite{Tchoumakov_2017} where the authors find that the degree of smoothness of an interface between topological materials does not interfere with the metallic interface modes and does, instead, introduce spurious massive interface modes in addition to the topological ones.

In order to find the interface states, firstly, we solve separately the two Schr\"{o}dinger equations associated with each half-space, then we impose the normalizability of the wavefunction, and its continuity at the interface at $z=0$. 
Since we focus on the interface at $z=0$, we replace $p_z$ with  $-i\partial_z$ while, because of the translational invariance along the $x$ and $y$ directions, $p_x$ and $p_y$ remain good quantum numbers.
The stationary Schr\"{o}dinger equation associated with each half-space is expressed as 
\begin{equation}\label{eq:S_Equations}
	\mathcal{H}_j \left(\bm{p}_{\perp},p_z \to -i\partial_{z}\right) \psi_j \left(\bm{r}\right) = \mathcal{E}\left(\bm{p}_\perp \right) \psi_j \left(\bm{r}\right),
\end{equation}
where $\bm{p}_{\perp}=\left(p_x,p_y\right)$, $\mathcal{E}\left(\bm{p}_\perp \right)$ is the eigenenergy, and all lengths are measured in units of $1/b$. 
To solve the problem above, we use the following ansatz~\cite{Zhang_2016_a}
\begin{equation}\label{eq:Ansatz}
	\psi_j (\bm{r}) = e^{ip_x x} e^{i p_y y} \begin{pmatrix}	\psi_1^j \\ \psi_2^j	\end{pmatrix} e^{\mu_j z},
\end{equation}
which is spatially localized close to the plane $z=0$ only if ${\rm Re}\left(\mu_j z\right) <0$, where $| {\rm Re} \left( \mu_j\right) |$ represents the localization length of the interface states around the $z=0$ plane.
By replacing Eq.~\eqref{eq:Ansatz} into Eq.~\eqref{eq:S_Equations}, we obtain the secular equation for the eigenenergies
\begin{equation}\label{eq:Secular_E}
	\det{\left[ \mathcal{H}_j \left(\bm{p}_\perp,\partial_z \to \mu_j\right) -\mathcal{E} \openone\right]} =0,
\end{equation}
which is solved by $\mu_j=\pm \bar{\mu}_j$, where
\begin{equation}\label{eq:Exponents}
	\bar{\mu}_j \equiv \frac{1}{u} \sqrt{\left[\bar{p}_j^2 -\left(p_x^2+p_y^2\right)\right]^2+\left(\lambda_j p_y\right)^2 -\mathcal{E}^2},
\end{equation}
which can be either a real or a pure imaginary number.
For each half-space $j$, by setting $\mu_j=\pm \bar{\mu}_j$, the non-trivial solutions of the homogeneous linear system $\left[\mathcal{H}\left(\bm{p}_\perp,\pm \bar{\mu}_j\right)-\mathcal{E}\right]\left(\psi_{1,\pm}^{j},\psi_{2,\pm}^j\right)^T=0$ are expressed as
\begin{equation}\label{eq:Spinors}
	\begin{pmatrix}	\psi_{1,\pm}^{j} \\ \psi_{2,\pm}^{j}	\end{pmatrix} = \begin{pmatrix}	\bar{p}_j^2 -\left(p_x^2 + p_y^2\right) - i\lambda_j p_y \\ \pm i u \bar{\mu}_j + \mathcal{E}	\end{pmatrix}.
\end{equation}
For a given $\bm{p}_\perp$, the general wavefunction is expressed in the spinorial form as
\begin{align}\label{eq:Wavefunction}
	\Psi_{\bm{p}_\perp} \left(\bm{r}\right) &=\mathcal{N}e^{i p_x x} e^{i p_y y} [\Theta(-z) \Phi_{\rm L} \left(z\right) +\Theta(z) \Phi_{\rm U} \left(z\right)],\\
	\Phi_j \left(z\right) &=  \sum_{\ell=\pm} C^j_{\ell} \begin{pmatrix}	\psi_{1,\ell}^j  \\ \psi_{2,\ell}^j 	\end{pmatrix} e^{\ell \bar{\mu}_j z},
\end{align}
where  $j\in \{\rm L,U\}$, ${\cal N}$ is the normalization prefactor, and $\Theta(z)$ is the Heaviside step function. 
The coefficients $\{C^j_{\ell} \}$ are determined by imposing the boundary conditions.
The first condition is the the normalizability of the wavefunction, which is equivalent to impose $\Psi (\bm{r}) \to 0$ for $|z|\to\infty$. The second condition is the continuity of the wavefunction at $z=0$. 
The first condition is satisfied by setting $C_{+}^{\rm U}=0$ and $ C_{-}^{\rm L}=0$.
Then, the continuity condition can be compactly expressed as 
\begin{equation}\label{eq:Continuity_Matrix}
{\cal M}
{\bm C}
=0,
\end{equation}
where
\begin{equation}\label{eq:Continuity_Matrix_Expression}
{\cal M}=
 \begin{pmatrix}	\psi_{1,-}^{\rm U} 	&-\psi_{1,+}^L \\ \psi_{2,-}^{\rm U} 	& -\psi_{2,+}^{\rm L }\end{pmatrix}~,
\end{equation}
and ${\bm C}=( C_{-}^{\rm U},C_{+}^{\rm L})^{\rm T}$.
For a given $\bm{p}_\perp$, Eq.~\eqref{eq:Continuity_Matrix} is solved by a non trivial set of 
coefficients $\{C_{-}^{\rm U}, C_{+}^{\rm U} \}$, 
for the values of energy ${\cal E}$ which nullify the determinant of ${\cal M}$.

\subsection{Interface electronic band}
\label{sec:interface_energy}

The secular equation $\det {\cal M}=0$ is explicitly expressed as
\begin{equation}\label{eq:Continuity_Eq}
	\begin{aligned}
	& \left(-i u \bar{\mu}_{\rm U} + \mathcal{E}\right)\left(\alpha_{\rm L} - i \lambda_{\rm L} p_y\right) \\
	&- \left(i u \bar{\mu}_{\rm L} + \mathcal{E}\right) \left(\alpha_{\rm U} - i \lambda_{\rm U} p_y\right) = 0,
	\end{aligned}
\end{equation}
where $\alpha_j = \bar{p}_j^2 - \left(p_x^2+p_y^2\right)$, and it is solved by
\begin{widetext}
\begin{equation}\label{eq:Solution_Energy_Light}
	\mathcal{E}\left(\bm{p}_\perp \right)= \frac{p_y \left\{ \left[ 1-\left(p_x^2 + p_y^2 \right)\right] \left( \Lambda_{\rm L}^2 \cos{\phi_{\rm L}} - \Lambda_{\rm U}^2 \cos{\phi_{\rm U}}\right) - \frac{e^2}{2} \Lambda_{\rm U}^2 \Lambda_{\rm L}^2 \left(\cos{\phi_{\rm L}} - \cos{\phi_{\rm U}}\right) \right\}}{\sqrt{\left( \frac{\Omega}{4 u}\right)^2 \left(\Lambda_{\rm U}^2-\Lambda_{\rm L}^2\right)^2 + \left(\Lambda_{\rm U}^2 \cos{\phi_{\rm U}} - \Lambda_{\rm L}^2 \cos{\phi_{\rm L}}\right)^2 p_y^2}},
\end{equation}
\end{widetext}
where the inequality $\cos\phi_{\rm U} \cos\phi_{\rm L}<0$ guarantees a mismatch in topology between the two regions and, consequently, the existence of the interface states.
For each $\bm{p}_\perp$, ${\cal E}(\bm{p}_\perp)$ represents the eigenenergy of an interface state if one has both ${\rm Re}\bar{\mu}_{\rm L}\neq 0$ and ${\rm Re}\bar{\mu}_{\rm U}\neq 0$. 
Otherwise, if ${\rm Re}\bar{\mu}_{\rm L}=0$ (${\rm Re} \bar{\mu}_{\rm U}=0$) the solution obtained is delocalized along the lower (upper) half-space, and it describes a bulk state.
Therefore, the condition $\bar{\mu}_{\rm L}\bar{\mu}_{ \rm U} = 0$  allows defining the boundaries of the domain of existence in the two-dimensional (2D) momentum space of the interface states.
We infer that the topological properties of the interface states depend on the relative arrangement in 2D momentum space $k_x-k_y$ of the Weyl points surface projections (WPPs) of the upper and lower FWSMs. 
In our system, this relative arrangement is determined by the different light intensities on the two half-spaces of the NLSM. Indeed, for each FWSM, through Eq.~\eqref{eq:H_FWSM_Factors_a}, the WPPs can be moved closer or farther away from the origin of momenta by changing the light intensity $\Lambda_j$. 
This degree of freedom can produce two relevant arrangements.
The symmetric case, using two identical light intensities $\Lambda_{\rm U}=\Lambda_{\rm L}$, where the WPPs of the two FWSMs are coincident, $\bar{p}_{\rm U}=\bar{p}_{\rm L}$, and the asymmetric case, using different light intensities $\Lambda_{\rm U}\neq \Lambda_{\rm L}$, 
where the WPPs of the two FWSMs are separated, $\bar{p}_{\rm U}\neq \bar{p}_{\rm L}$.
Figure~\ref{fig:EnergyDispersionNonMagnetic} shows the density plots of the energy dispersion of the interface states as a function of the momentum components $p_x$ and $p_y$, in the symmetric case, in panel~(a), and the asymmetric case, in panel~(b).
In Fig.~\ref{fig:EnergyDispersionNonMagnetic}~(a), where $\Lambda_{\rm U}=\Lambda_{\rm L}=3$, the interface states are well defined in the whole $2$D momentum space with the exception of the $p_y=0$ axis (solid white line). 
Along the $p_y=0$ axis, for each $p_x$ the electronic band composed of the interface states merges into the bulk conduction (valence) band at the energy $E=+|\bar{p}^2-p_x^2|$ ($E=-|\bar{p}^2-p_x^2|$).
In Fig.~\ref{fig:EnergyDispersionNonMagnetic}~(b), where $\Lambda_{\rm U}=2$ and $\Lambda_{\rm L}=3$, the interface states are delimited by solid (dashed) boundary lines, which are the solutions of $\bar{\mu}_{\rm U}=0$  ($\bar{\mu}_{\rm L}=0$). In particular, at the red (blue) boundary lines the interface band merges into a conduction (valence) bulk band. 
The Fermi line, black line in Fig.~\ref{fig:EnergyDispersionNonMagnetic}~(b), is composed of a circumference that surrounds the origin of momenta and two segments that lay along the $ p_y = 0 $ axis. The intersections of the segments and the circumference of the Fermi line correspond to two saddle points.
The appearance of the saddle points at the Fermi energy leads to a Van Hove singularity in the density of interface states (DOS)~\cite{Grosso_2000}.
Thus, starting from the dispersion relation of the interface states, we write the corresponding DOS as
\begin{align}\label{eq:DOS_general}
	\rho \left(\mathcal{E}\right) &= \int \frac{d \bm{p}_\perp }{\left(2\pi\right)^2} \delta \left(\mathcal{E} - \mathcal{E}\left(\bm{p}_\perp \right)\right),
\end{align}
where $\delta(x)$ is the Dirac delta function.
\begin{figure}[t!] 
\centering
\begin{overpic}[width=0.9\columnwidth]{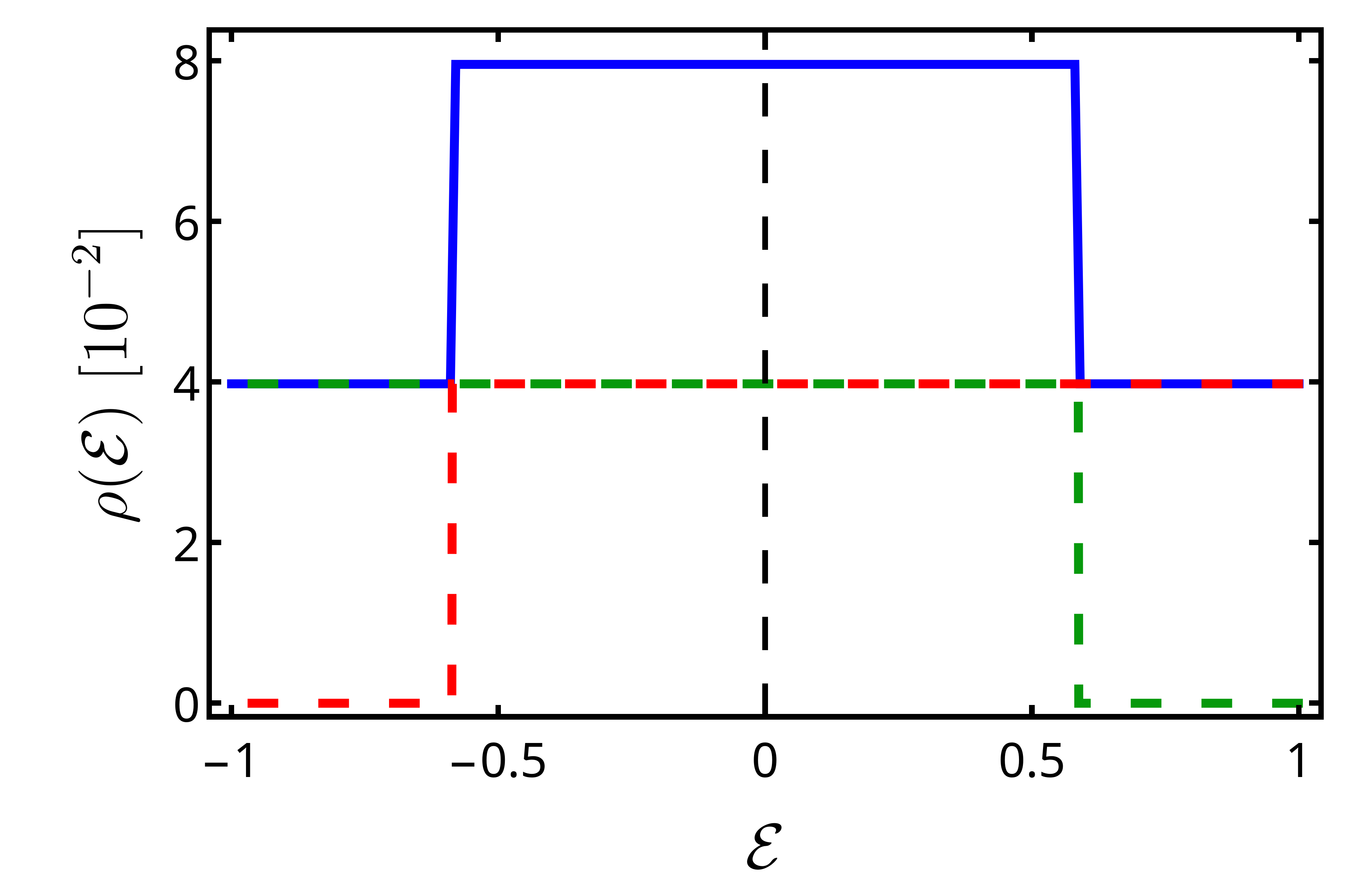}\put(0,65){(a)}\end{overpic}
\centering
\begin{overpic}[width=0.9\columnwidth]{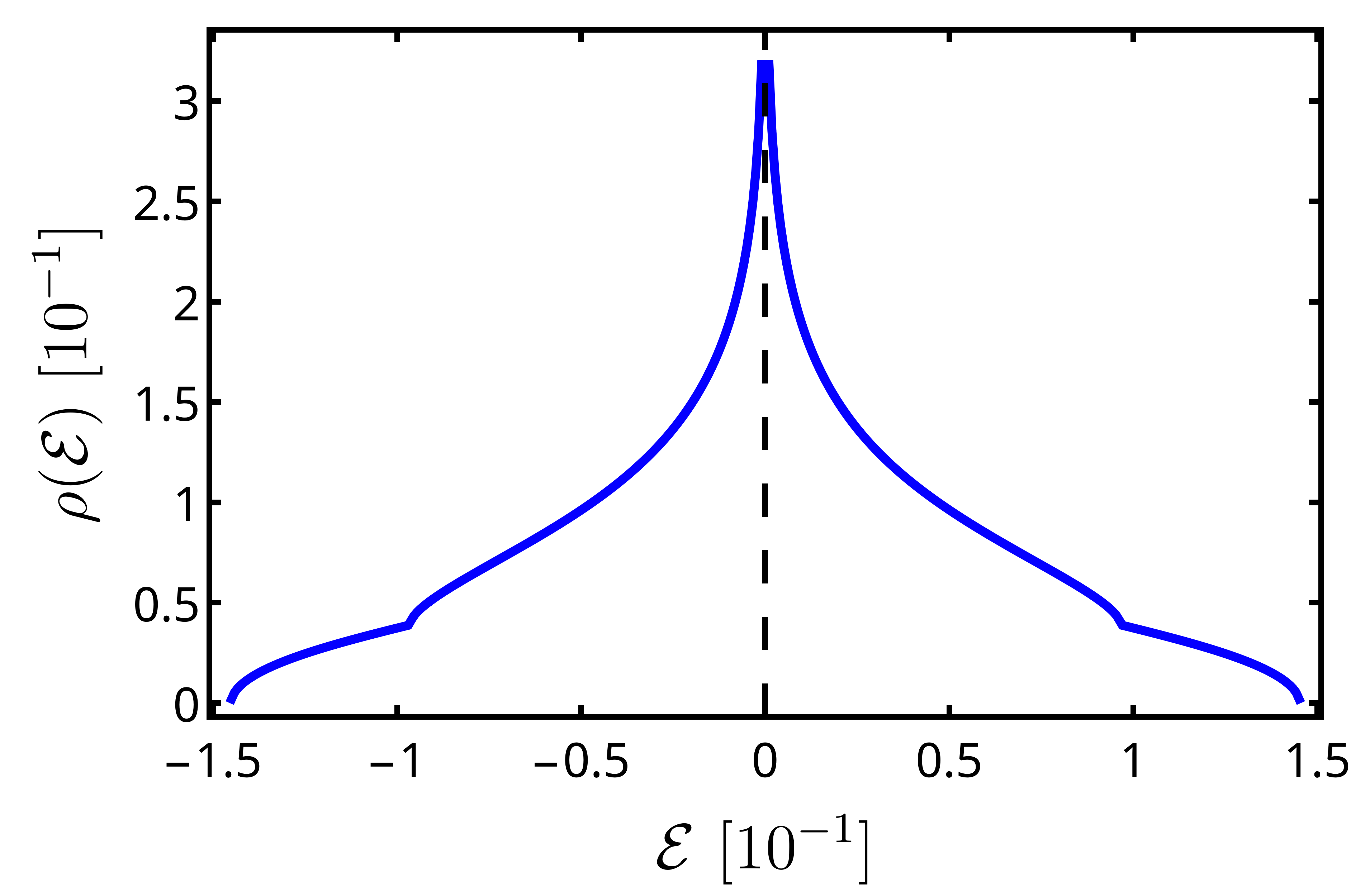}\put(0,65){(b)}\end{overpic}
  \caption{Density of interface states as a function of the energy $\cal E$, shown in the blue line in both panels. 
  Panels (a) refers to the symmetric case ($A_{\rm U}=A_{\rm L}=3$), where the supplemental dashed red [green] line corresponds to the contribution
   $\Theta \left( \mathcal{E}+\bar{p}^2 \right)/(8\pi)$ [$\Theta\left(-\mathcal{E}+\bar{p}^2\right)/(8\pi)$].
 Panel (b) refers to the asymmetric case ($A_{\rm U}=2$, and $A_{\rm L}=3$), where at zero energy a Van Hove singularity occurs. 
 In all panels, we set $u=1,\Omega=10$, $\phi_{\rm U} = \pi$ and $\phi_{\rm L}=0$.}
  \label{fig:DOSNonMagnetic}
\end{figure}
In the symmetric case, setting the opposite polarizations $\phi_{\rm U} = \pi$ and $\phi_{\rm L} = 0$, the energy dispersion has the following quadratic form
\begin{equation}\label{eq:Energy_Symmetric}
\mathcal{E} \left( \bm{p}_{\perp}\right) = \sign{\left(p_y\right)} \left[\bar{p}^2 - \left| \bm{p}_{\perp} \right|^2 \right],
\end{equation}
and it is straightforward to obtain an analytical expression for the DOS
\begin{equation}\label{eq:DOS_Symmetric}
\rho({\cal E}) = \frac{1}{8\pi}\left[ \Theta \left( -\mathcal{E} + \bar{p}^2 \right) + \Theta \left( \mathcal{E} + \bar{p}^2 \right) \right],
\end{equation}
that is shown in Fig.~\ref{fig:DOSNonMagnetic}~(a).
Figure~\ref{fig:DOSNonMagnetic}~(b) displays the DOS in the asymmetric case, where a Van Hove singularity appears at the Fermi energy ${\cal E}=0$. A Van Hove singularity always appears in the DOS when $\Lambda_{\rm U}\neq\Lambda_{\rm L}$, independently of the specific values of $\Lambda_{\rm U}$ and $\Lambda_{\rm L}$.
Therefore, we have seen that by tuning the intensities of the two beams of light, one can change the WPPs arrangements, generating modifications both in the domain of existence of the interface states and in the shape of the Fermi line. 
This is the first most important result of this work: one can engineer a Van Hove singularity in a 2D interface electron system by illuminating a NLSM with two beams of lights of different intensities and opposite polarizations.

\subsection{Pseudo-spin texture}
\label{sec:interface_pseudospin}

\begin{figure}[t] 
\begin{overpic}[height=6.35cm,trim={0 0cm 0 -0.25cm}]{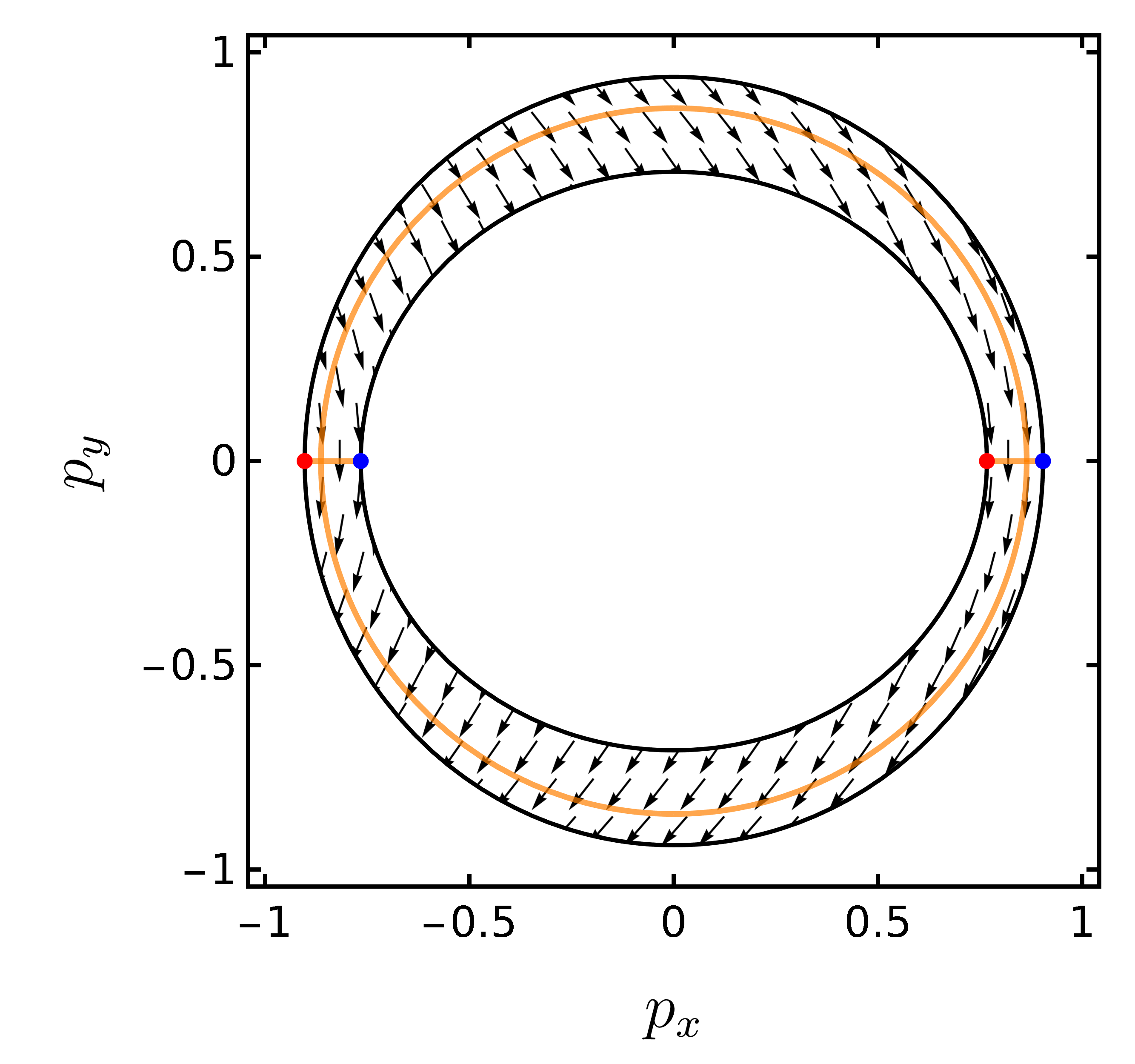}\end{overpic}
\caption{Pseudo-spin texture of the interface states in the asymmetric case ($\Lambda_{\rm U}\neq \Lambda_{\rm L}$), which shows a trivial structure with zero winding number.
The orange solid line represents the Fermi line ${\cal E}=0$, and the red and blue dots denote the WPPs of positive and negative chirality, respectively. The parameters used are: $u=1,\Omega=10$, $\Lambda_{\rm U}=2, \Lambda_{\rm L}=3$, $\phi_{\rm U} = \pi$, and $\phi_{\rm L}=0$.
  \label{fig:PseudoSpinNonSymmetric}}
\end{figure}

Next, we analyze the pseudo-spin texture of the interface eigenstates, which gives further information on the topological nature of the interface system.
For each interface eigenstate, labeled by  $\bm{p}_\perp$, we define the corresponding pseudo-spin vector as 
\begin{equation}\label{eq:Pseudospin_definition}
	\left\langle \bm{\sigma}\right\rangle_{\bm{p}_\perp} = 
	\int d\bm{r} \Psi_{\bm{p}_\perp}^\dagger(\bm{r}) \bm{\sigma} \Psi_{\bm{p}_\perp}(\bm{r}),
\end{equation}
where $\Psi_{\bm{p}_\perp}(\bm{r})$ is the wavefunction expressed in spinorial form, accordingly to Eq.~\eqref{eq:Wavefunction}.
In the symmetric case ($A_{\rm U}=A_{\rm L}$) the pseudo-spin texture has a trivial pattern. Here, the pseudo-spins are all aligned along the $x$-direction, i.e. $\langle \bm{\sigma} \rangle_{\bm{p}_\perp} = \left( \sign{\left(\lambda p_y\right)},0,0\right)$.
In the asymmetric case  ($A_{\rm U}\neq A_{\rm L}$) the domain of existence of the interface states is homeomorphic to an annulus (see Fig.~\ref{fig:EnergyDispersionNonMagnetic} (b)),
and we characterize the topological properties of the interface system by focusing on the generic closed paths that cannot be shrunk into points.
In particular, we calculate the winding number~\cite{Asboth_2016}, which is defined for any closed path $\Gamma$ parametrized by $\tau \in \left[0,1\right]$ as 
\begin{equation}\label{eq:WindingNumber_Definition}
\nu\left(\Gamma\right) = \frac{1}{2\pi} \int_{0}^{1} \left( \left\langle \bm{\sigma} \right\rangle_{\bm{p}_\perp} \left( \tau \right) \times \frac{d}{d\tau} \left\langle \bm{\sigma} \right\rangle_{\bm{p}_\perp} \left( \tau \right) \right)_z d\tau,
\end{equation}
where $\left\langle \bm{\sigma}\right\rangle_{\bm{p}_\perp} \left(\tau\right)$ is the interface states' pseudo-spin of Eq.~\eqref{eq:Pseudospin_definition} calculated at the $\bm{p}_\perp$ corresponding to the parametric variable $\tau$ along the curve $\Gamma$. 
A nonzero integer $\nu$ corresponds to $|\nu|$-complete rotations of the pseudo-spin, namely a topologically nontrivial pseudo-spin texture.
Figure~\ref{fig:PseudoSpinNonSymmetric} shows the pseudo-spin pattern within the domain of existence of the interface states in the asymmetric case, and the orange solid line represents the Fermi line ${\cal E}(\bm{p}_\perp)=0$.
Along the circumference at zero energy shown in Fig.~\ref{fig:PseudoSpinNonSymmetric} and parametrized by $(p_x,p_y)= R (\cos(\theta),\sin(\theta))$, where $R=\sqrt{\left(\lambda_U \bar{p}^{2}_L - \lambda_L \bar{p}^{2}_U\right) / \left(\lambda_U -\lambda_L\right)}$, we obtain
\begin{equation}\label{eq:Circular_Arc_Pseudospin}
	\left\langle \bm{\sigma} \right\rangle_{\bm{p}_\perp} \propto \left( \lambda_{\rm U} R \sin{\theta}, R^2 - \bar{p}_{\rm U}^2,0 \right).
\end{equation}
Despite a nonzero $y$-component occurring, along this circular path, the pseudo-spin texture does not make a full rotation, and the winding number is zero. Hence, independently of values of the light intensities, the pseudo-spin texture is topologically trivial.

\section{Magnetic barrier}\label{sec:interface_magnetic}

\begin{figure}[t] 
\centering
\begin{overpic}[height=6.45cm]{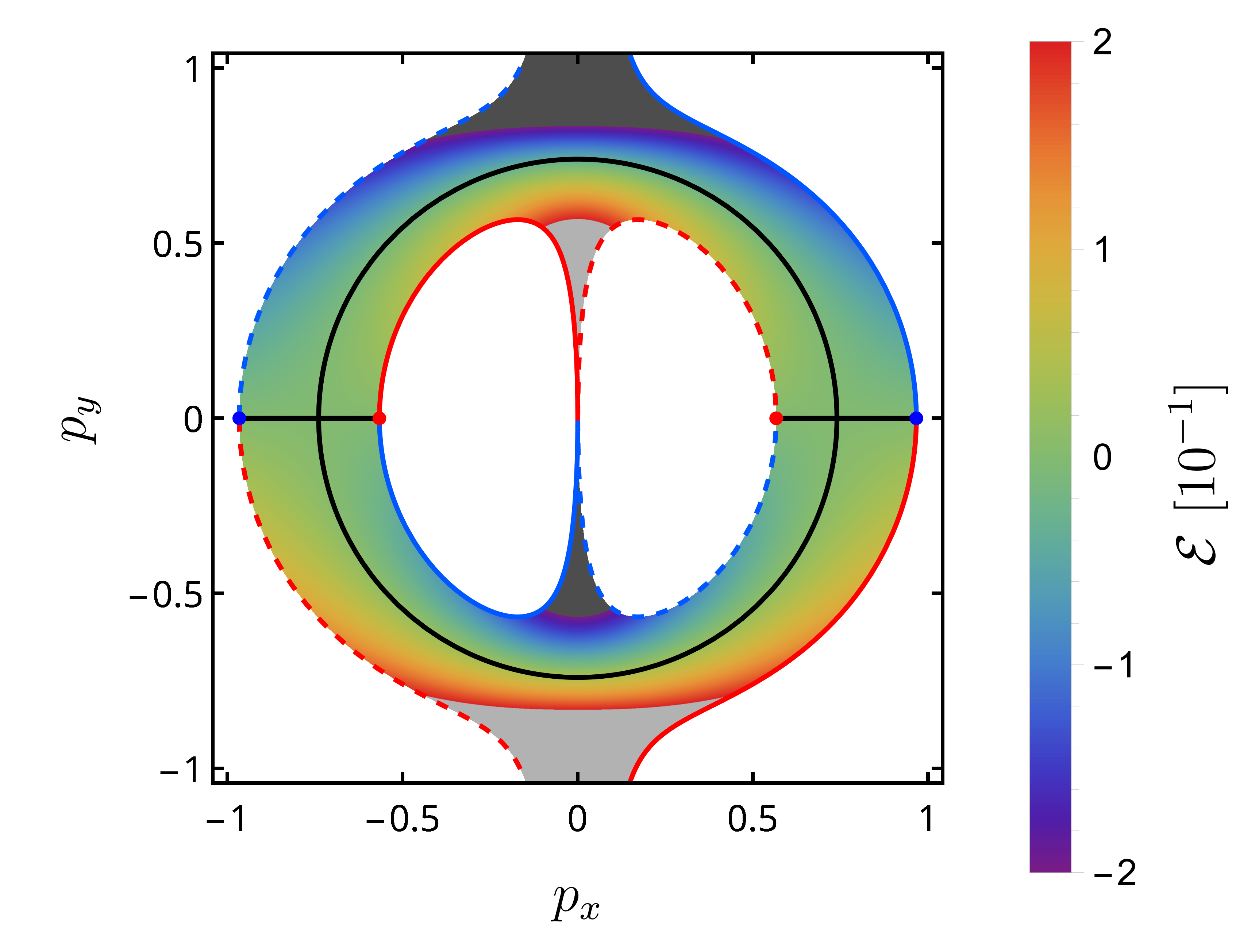}\put(4,75){(a)}\end{overpic}\\
\hspace{-6.05em}\begin{overpic}[height=6.20cm]{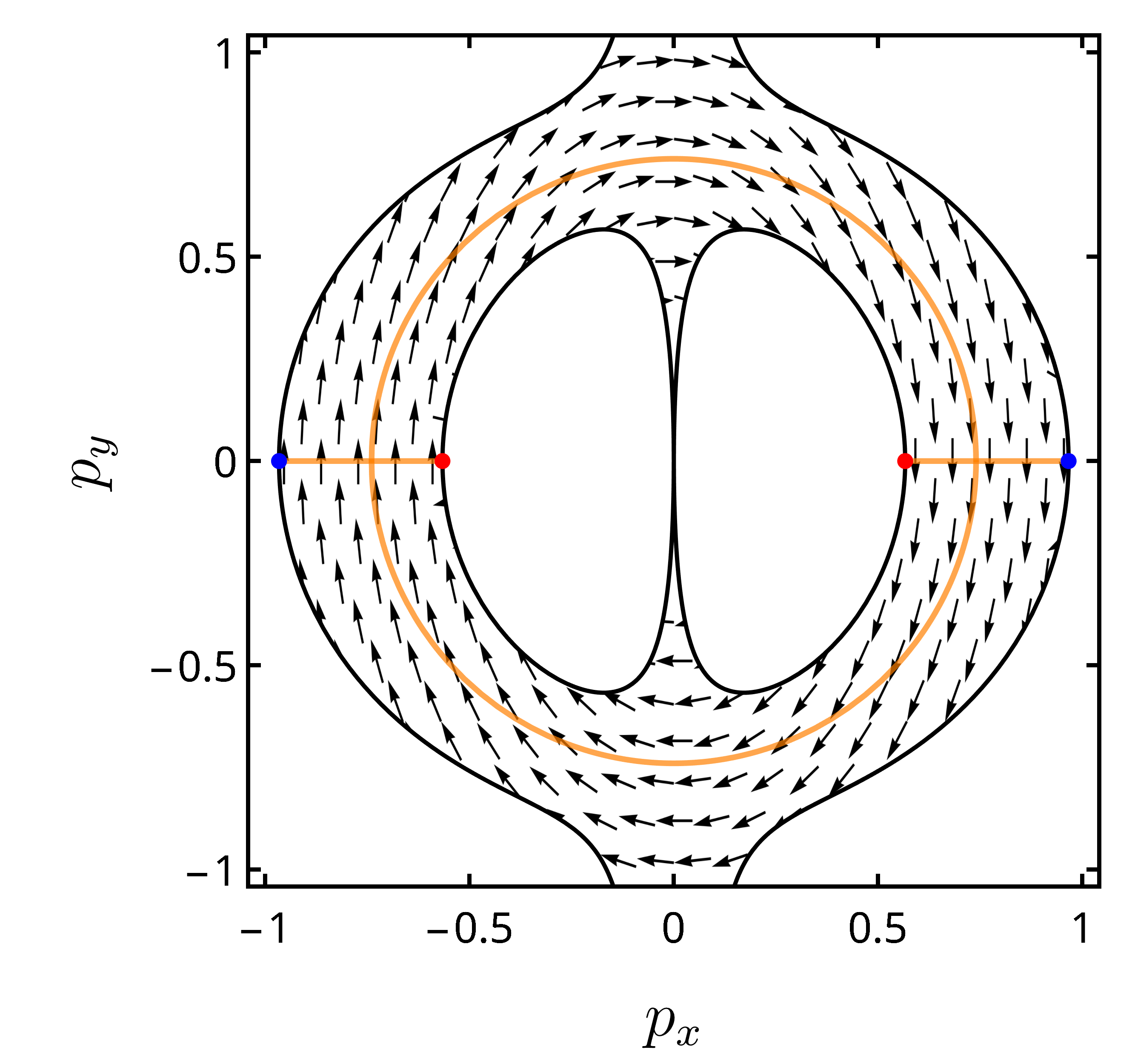}\put(4,95){(b)}\end{overpic}
  \caption{(a) Density plot of the energy dispersion, and (b) pseudo-spin texture plot of the interface states as a function of the momentum components $p_x$ and $p_y$,  with $\Lambda_{\rm U}=\Lambda_{\rm L}$ and in the presence of a delta-like magnetic field at $z=0$.
 In panel (a), the boundary solid (dashed) lines represent the merging of the interface states' band with the upper (lower) FWSM bulk eigenbands. The red (blue) lines describe the merging into the conduction (valence) band, and red and blue dots represent WPPs of positive and negative chirality, respectively. Positive (negative) out of range values are depicted as light (dark) gray areas, and the Fermi line $\mathcal{E}=0$ is shown in black in panel (a) and in orange in panel (b). The parameters used are: $u=1,\Omega=10$, $\Lambda_{\rm U}=\Lambda_{\rm L}=3$, $\phi_{\rm U} = \pi$, $ \phi_{\rm L}=0$, and the magnetic parameter is set at $p_0 = 1/5$.
  \label{fig:Magnetic}}
\end{figure}

In this section, we introduce a delta-like magnetic barrier along the interface between the two FWSM regions, and we analyze how this additional term can induce a non-trivial pseudo-spin texture.
Firstly, we study the modifications in the interface states of the setup in Fig.~\ref{fig:setup} by adding the following magnetic barrier~\cite{Wu_2010}
\begin{equation}\label{eq:MagneticBarrier}
	\bm{B}\left(\bm{r}\right)=B_0 \delta(z) \hat{y},
\end{equation}
where we remind that all lengths are measured in units of $1/b$.
Within our formalism, we introduce the effect of the magnetic field by resorting to the Peierls substitution $\bm{p}\to\bm{p}+e\bm{A}_0/b$, 
where the vector potential $\bm{A_0}$ generates the magnetic field through the definition $\bm{B} = b (\partial/\partial_{\bm{r}}) \times \bm{A}_0$, and it is expressed as
\begin{equation}\label{eq:BarrierPotential}
	\bm{A}_0\left(\bm{r}\right) = \frac{B_0}{2 b}\sign{\left(z\right)}\hat{x}.
\end{equation}
For each half-space, the FWSM Hamiltonian of Eq.~\eqref{eq:H_Setup} is rewritten as
\begin{align}
	\mathcal{H}_j \left(\bm{p}\right) & = \left\{\bar{p}_j^2 - \left[\left(p_x - \zeta_j p_0\right)^2 + p_y^2\right]\right\} \sigma_x \nonumber \\ &+ \lambda_j p_y \sigma_y + u p_z \sigma_z, \label{eq:H_Setup_Magnetic}
\end{align}
where $j\in\{\rm L,U\}$, $p_0=- e B_0/(2 b^2)$, $\bar{p}_j$ and $\lambda_j$ are defined in Eqs.~\eqref{eq:H_FWSM_Factors_LU}, and $\zeta_{\rm U}=+1$ and $\zeta_{\rm L}=-1$.
In each FWSM the WPs are placed at $\bm{P}_{j,\pm} = \left(\pm \bar{p}_j + \zeta_j p_0,0,0\right)$. 
The magnetic term $p_0$ induces a rigid shift of each pair of WPs along the $x$-direction of the 2D momentum space.
Hence, the intensity of the delta-like magnetic field, parametrized by $p_0$, represents a further knob for modifying the WPPs of the FWSMs,
in addition to the intensities and polarizations of the beams of light.
Here, we see how this further degree of freedom can enrich the energy dispersion and alter the structure of the pseudo-spin pattern.
For sake of simplicity, we consider the symmetric case ($\Lambda_{\rm U}=\Lambda_{\rm L}=\Lambda$, $\phi_{\rm U}=\pi$, and $\phi_{\rm L}=0$), where the dispersion relation assumes the following compact expression
\begin{equation}\label{eq:EnergyDispersion_Magnetic_Simplified}
	\mathcal{E}\left(\bm{p}_{\perp}\right) = p_y \Lambda^2 \frac{1-\frac{e^2 \Lambda^2}{2}-\left(p_x^2 + p_y^2 + p_0^2\right)}{\sqrt{\left(\frac{\Omega}{u}\right)^2 \left(p_x p_0\right)^2 +\left(\Lambda^2 p_y\right)^2}}.
\end{equation}
Figure~\ref{fig:Magnetic}~(a) shows the density plot of the dispersion relation for this symmetric case. 
Like in the asymmetric case without magnetic barrier, there are two saddle points in the Fermi line (black line in Fig.~\ref{fig:Magnetic}~(a)) which cause the appearance of a Van Hove singularity in the DOS.
Moreover, in Fig.~\ref{fig:Magnetic}~(a), the solid (dashed) boundary lines are solutions of the delocalization condition $\bar{\mu}_{\rm U}=0$ ($\bar{\mu}_{\rm L}=0$), where
\begin{equation}
\bar{\mu}_j=\frac{1}{u} \sqrt{\left\{\bar{p}_j^2 -\left[(p_x-\zeta_j p_0)^2+p_y^2\right]\right\}^2+\left(\lambda_j p_y\right)^2 -\mathcal{E}^2}, 
\end{equation}
and red (blue) lines describe the merging of the band composed by the interface states into a conduction (valence) bulk band.
Besides a modification of the energy dispersion relation, the presence of the delta-like magnetic field has a strong impact on the pseudo-spin texture. 
Figure~\ref{fig:Magnetic}~(b) displays the pseudo-spin pattern within the domain of existence of the interface states in the symmetric case, and the orange solid line represents the Fermi line ${\cal E}(\bm{p}_\perp)=0$.
Along the circumference at zero energy shown in Fig~\ref{fig:Magnetic}~(b) and parametrized by $(p_x,p_y)=\bar{R}(\cos(\theta),\sin(\theta))$, where $\bar{R} = \sqrt{\bar{p}^2-p_0^2}$, we obtain 
\begin{equation}\label{eq:Circular_Arc_Pseudospin_Magnetic}
	\left\langle \bm{\sigma} \right\rangle_{\bm{p}_\perp} \propto \left(\lambda \sin\theta , -2 p_0 \cos\theta,0\right),
\end{equation}
where $\bar{p}$ and $\lambda$ are defined in Eqs.~\eqref{eq:H_FWSM_Factors}.
In this case, the pseudo-spin pattern has a nontrivial structure. In fact, along this closed path, the pseudo-spin makes a complete rotation, and  we find a non-vanishing winding number $\nu=+1$,  where the sign $+$  is given by the counterclockwise rotation of the pseudo-spin.
This is the second most important result of this work: the presence of a magnetic barrier along the interface between the two FWSM regions can induce a nontrivial topology in the pseudo-spin pattern.
Specifically, the pseudo-spin structure is nontrivial under the condition
\begin{equation}\label{eq:NonTrivialityCondition}
	\left| \bar{p}_{\rm U} - \bar{p}_{\rm L} \right| < 2\left| p_0 \right| < \bar{p}_{\rm U} + \bar{p}_{\rm L}~,
\end{equation}
where $\bar{p}_j$ is defined by Eq.~\eqref{eq:H_FWSM_Factors__LU_a}.
For any given magnetic parameter $|p_0|<1$, by choosing suitable values of the light intensities, it is possible to set the values of $\bar{p}_{\rm U}$ and $\bar{p}_{\rm L}$ such that they fulfill the conditions of Eq.~\eqref{eq:NonTrivialityCondition}. By conveniently tuning the light intensities, one can also set the values of $\bar{p}_{\rm U}$ and $\bar{p}_{\rm L}$ such that the inequalities in Eq.~\eqref{eq:NonTrivialityCondition} are not satisfied.
Hence, in the presence of the magnetic barrier, it is possible to switch the topological properties of the interface states by only modulating the light intensities. 
We note that the condition in Eq.~\eqref{eq:NonTrivialityCondition} can be interpreted in terms of the WPPs arrangement. In particular, both inequalities of Eq.~\eqref{eq:NonTrivialityCondition} are satisfied if the inner (outer) WPPs have identical chiralities, an example is shown in Fig.~\ref{fig:Magnetic}.
In the previous section, we have verified that this type of arrangement of the WPPs is not reachable by exploiting only the light beams, and for this reason, the assistance of the local magnetic field appears crucial.

%

\begin{figure*}[t!] 
\begin{minipage}{0.49\textwidth}
\begin{overpic}[height=6.45cm]{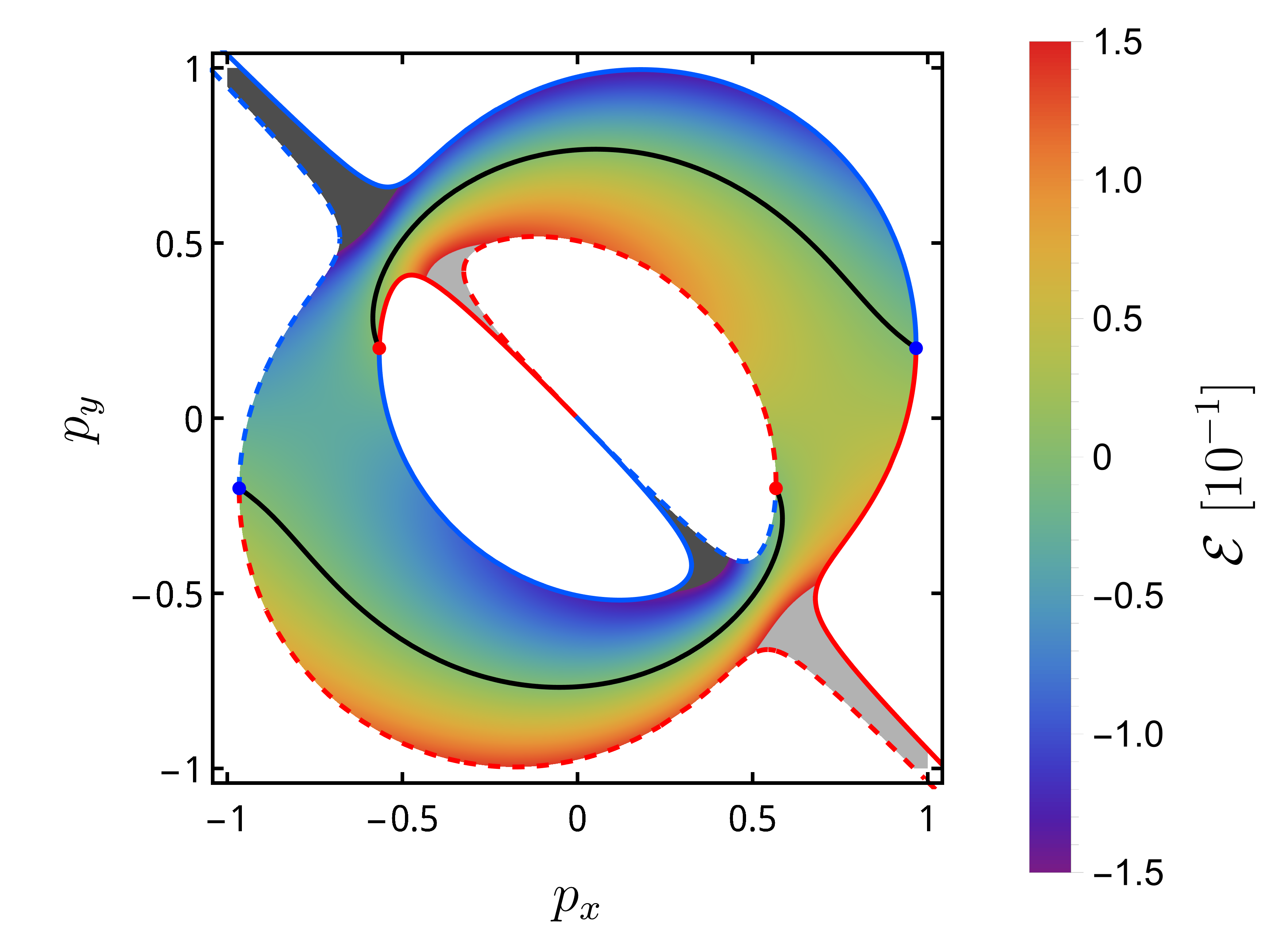}\put(4,75){(a)}\end{overpic}
\end{minipage}
\begin{minipage}{0.49\textwidth}
\vspace{.4em}
\begin{overpic}[height=6.20cm]{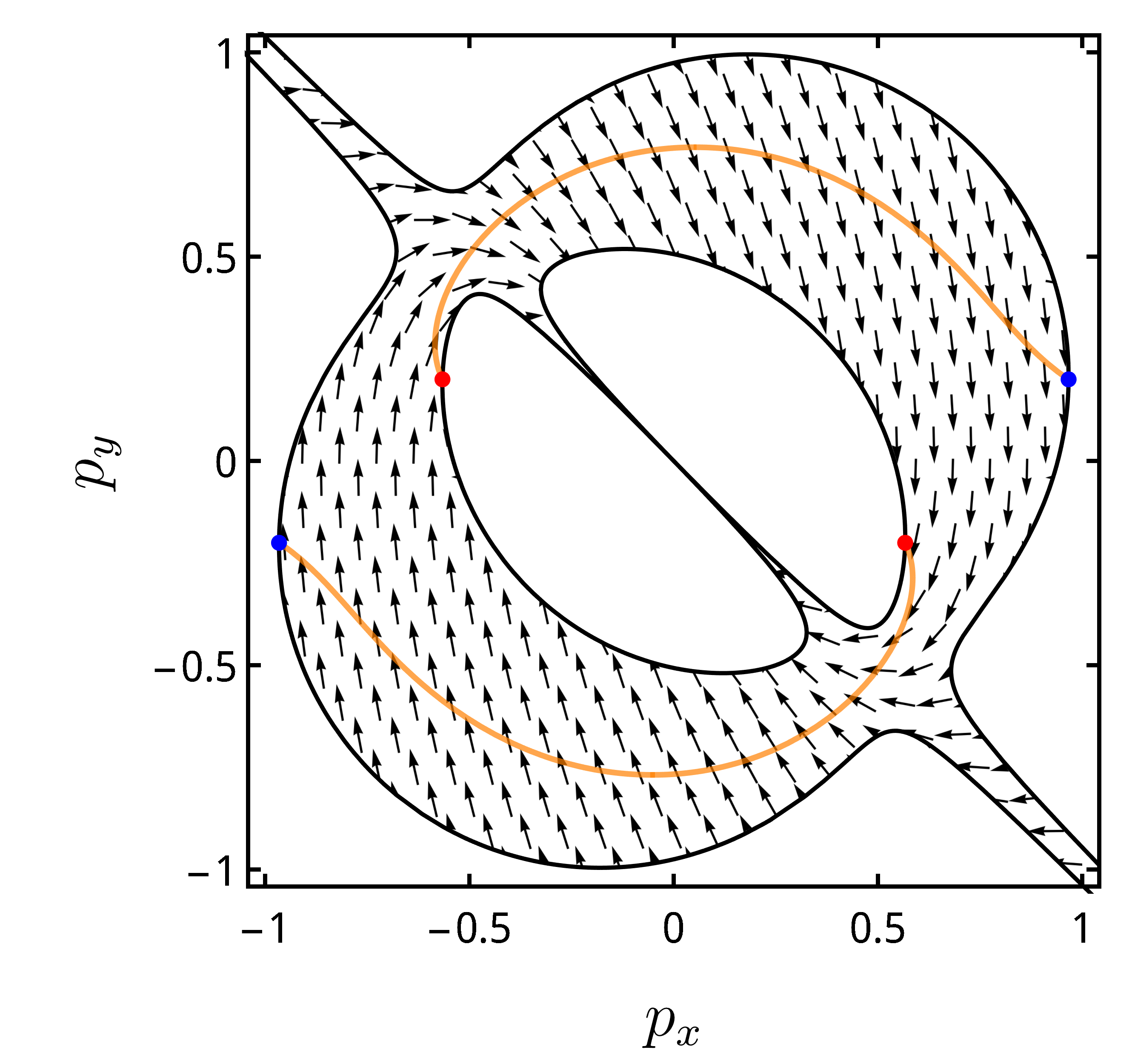}\put(4,97){(b)}\end{overpic}
\end{minipage}
  \caption{(a) Density plot of the energy dispersion, and (b) pseudo-spin texture plot of the interface states as a function of the momentum components $p_x$ and $p_y$,  with $\Lambda_{\rm U}=\Lambda_{\rm L}$ and in the presence of a delta-like magnetic field at $z=0$ with components both along the $x$ and $y$ directions.
 In panel (a), the boundary solid (dashed) lines represent the merging of the interface states' band with the upper (lower) FWSM bulk eigenbands. The red (blue) lines describe the merging into the conduction (valence) band, and red and blue dots represent WPPs of positive and negative chirality, respectively. Positive (negative) out of range values are depicted as light (dark) gray areas, and the Fermi lines $\mathcal{E}=0$ are shown in black in panel (a) and in orange in panel (b). The parameters used are: $u=1,\Omega=10$, $\Lambda_{\rm U}=\Lambda_{\rm L}=3$, $\phi_{\rm U} = \pi$, $ \phi_{\rm L}=0$, and the magnetic parameters are set at $p_{0,x} = 1/5$ and $p_{0,y} = 1/5$.
   \label{fig:MagneticNonSymmetric}}
\end{figure*}

Finally, we analyse the effects of a magnetic barrier which has a field component also along the $x$ direction. The magnetic barrier can be written as
\begin{equation}
	\bm{B}\left(\bm{r}\right) = \delta\left( z \right) \left( B_{0,x} \hat{x} + B_{0,y} \hat{y} \right)
\end{equation}
which, after the Peierls substitution, leads for each half-space $j \in \{\rm{L,U}\}$ to the FWSM Hamiltonian
\begin{align}\label{eq:Hamiltonian_FieldXY}
	\mathcal{H}_j \left(\bm{p}\right) & = \left\{\bar{p}_j^2 - \left[\left(p_x - \zeta_j p_{0,x}\right)^2 + \left(p_y-\zeta_j p_{0,y}\right)^2\right]\right\} \sigma_x \nonumber \\ &+ \lambda_j \left(p_y-\zeta_j p_{0,y} \right) \sigma_y + u p_z \sigma_z, 
\end{align}
 where $\left( p_{0,x}, p_{0,y} \right) = \left(-eB_{0,y}/(2b^2),eB_{0,x}/(2b^2)\right)$, $\zeta_U = +1$ and $\zeta_L = -1$, and $\bar{p}_j$ and $\lambda_j$ are still given by Eqs.~\eqref{eq:H_FWSM_Factors_LU}. Figures \ref{fig:MagneticNonSymmetric} (a) and \ref{fig:MagneticNonSymmetric} (b) show the energy dispersion and pseudo-spin texture, respectively, of the interface system, derived from the Hamiltonian Eq.~\eqref{eq:Hamiltonian_FieldXY} when the $x$ and $y$ component of the magnetic barrier are equal in strengths, $p_{0,x} = p_{0,y}$. The meanings of lines and dots is the same as in Fig.~\ref{fig:Magnetic}. It can be seen from Fig.~\ref{fig:MagneticNonSymmetric} (b) that the pseudo-spin texture is still non-trivial and all paths which enclose the origin of momenta will have a non vanishing winding number. For higher $x$ components of the magnetic barrier, roughly for $p_{0,y} \gtrsim 2 p_{0,x}$, the domain of existence of the interface states is no longer connected and we cannot properly define its topology through the winding number of Eq.~\eqref{eq:WindingNumber_Definition}. The DOS of the interface system is also slightly affected. In particular, the Van Hove singularity at the Fermi energy is split into two energy symmetric ones. The energies of these new Van Hove singularities is numerically found to be linearly proportional to the strength of the $x$ component of the magnetic field, $\mathcal{E}_{\rm VHS} \sim \pm p_{0,y}$ (up to a maximum value, with the given parameters, $p_{0,y} \gtrsim 3.5 p_{0,x}$, after which they no longer exist).

\section{Conclusions}

In this paper, we analyzed the boundary states that emerge at the interface between two sides of an infinite NLSM that are illuminated by monochromatic light beams of opposite circular polarizations. In particular, we focused on the energy dispersion, DOS, and pseudo-spin texture.

Illuminating the system with lights of opposite polarization generates two FWSMs with opposite Chern numbers, and this topology mismatch leads to the appearance of topological interface states.
We have shown that the topological properties of the interface states strictly depend to the relative arrangements of the WPPs of the two induced WSMs. 
The independent tunability of the light intensities represents a knob for modifying this relative arrangement. As such, by changing the intensities of the light beams one can modify the domain of existence of the interface states along the 2D reciprocal space, and the shape of the Fermi line. 
In particular, illuminating the system with two different light intensities induces a transition in the Fermi line with the creation of a Van Hove singularity in the DOS.

Moreover, we have added a further knob for modifying the interface states, namely we have introduced a magnetic barrier along the interface given, for instance, by localized doping with magnetic impurities. 
The presence of this local magnetic field together with the tunability of the light intensity allows to change the relative arrangement of WPPs at will. We have thus found that the $y$ component of this magnetic barrier can lead to the creation of a novel non-trivial pseudo-spin pattern of the interface states. The $x$ component, on the other hand, works against the non-trivial topology of the interface system; nonetheless, the system has a fair degree of robustness, roughly up to $B_x \lesssim 2 B_y$ in the setup studied. In this paper, we have classified the topology of the interface system in terms of the winding of the pseudo-spin around the origin of momenta, and we have shown that by modulating the light intensities it is possible to switch on/off the non-trivial topology.
	
Together with the presence of VHSs close to the Fermi level, the discovered tunable topological interface state could provide promising applications in the search for exotic correlated quantum phases of matter and for optoelectronics.

\acknowledgements

The authors thank G. G. N. Angilella, E. Paladino, and B. Trauzettel for illuminating discussions and fruitful comments on various stages of this work.
This research was supported by the Universit\`a degli Studi di Catania, Piano di Incentivi per la Ricerca di Ateneo 2020/2022 (progetto QUAPHENE and progetto Q-ICT).
F.B. thanks the University of W\"{u}rzburg for its hospitality.

\bibliographystyle{mprsty}
\bibliography{Bibliography}

\appendix
\numberwithin{equation}{section}
\renewcommand\thefigure{\thesection.\arabic{figure}}
\setcounter{figure}{0}

\section{Numerical results}\label{sec:numerical}
\begin{figure}[t!] 
\centering
\begin{overpic}[height=6.45cm]{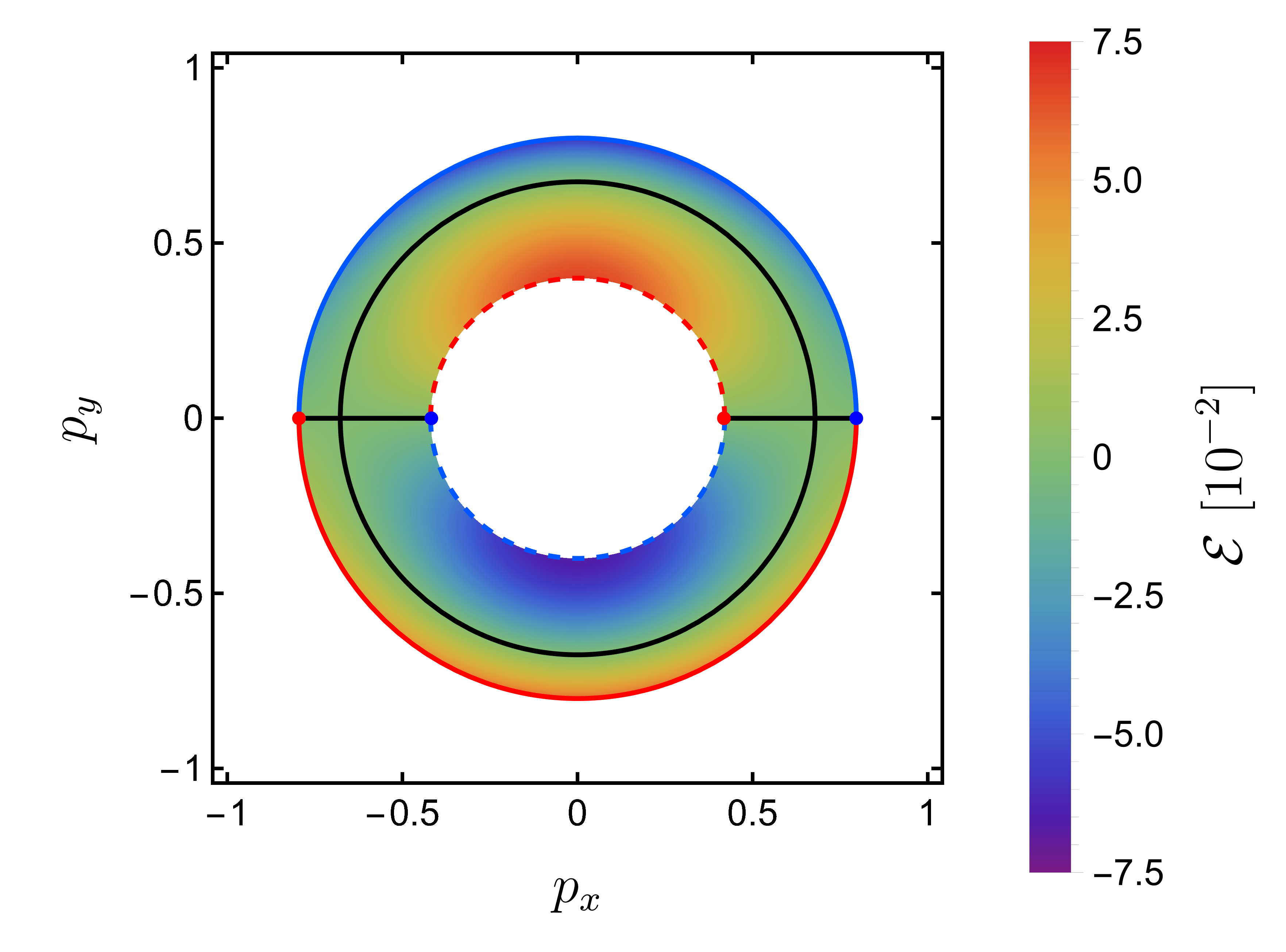}\put(4,75){(a)}\end{overpic}\\
\hspace{-6.30em}\begin{overpic}[height=6.15cm]{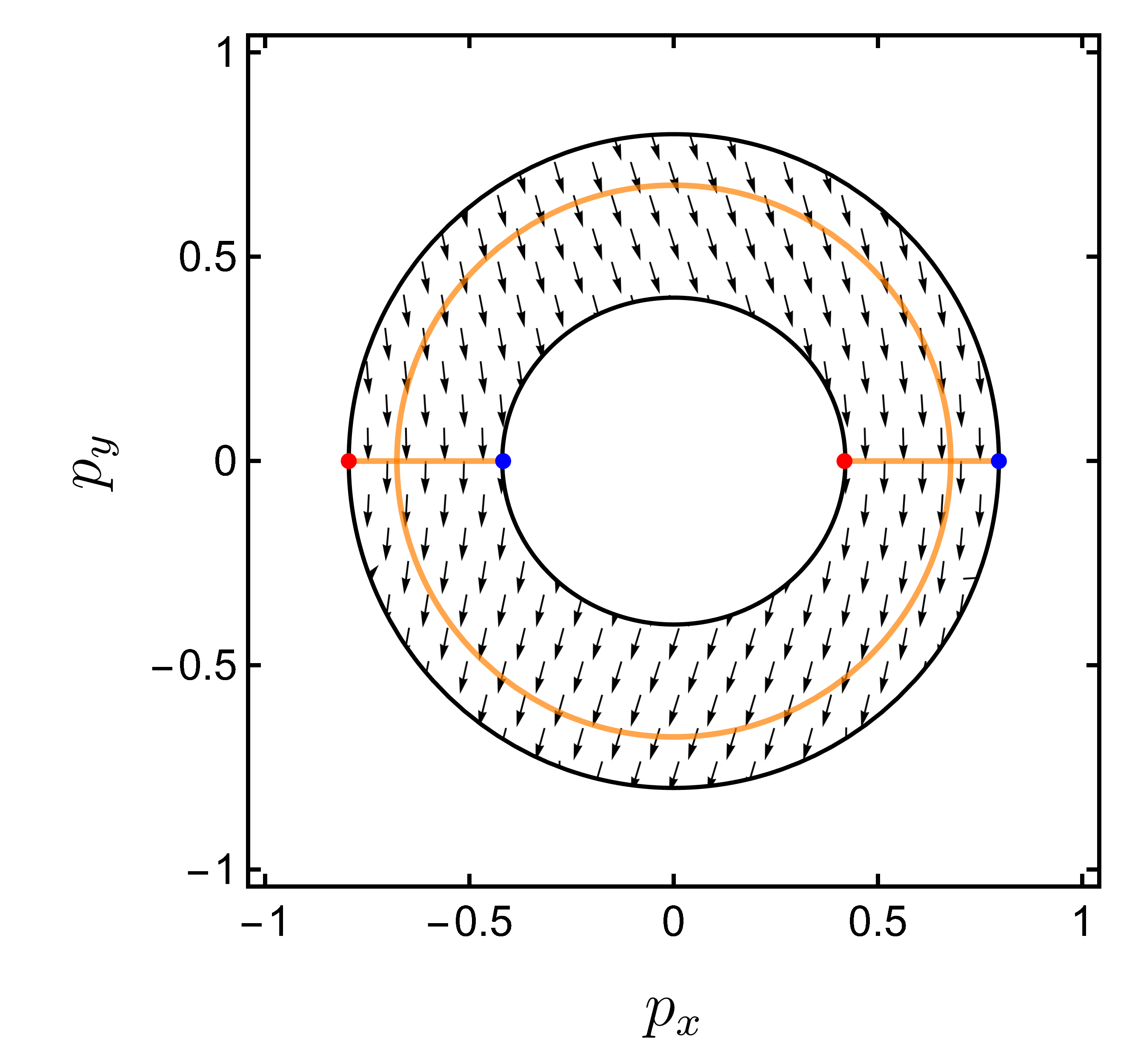}\put(4,95){(b)}\end{overpic}
  \caption{(a) Density plot of the energy dispersion, and (b) pseudo-spin texture plot of the interface states as a function of the momentum components $p_x$ and $p_y$ in the asymmetric case ($\Lambda_{\rm U}<\Lambda_{\rm L}$), by holding the quadratic term in $p_z$ in the Hamiltonian~\eqref{eq:H_NLSM_Dimensionless_Numerical} which describes the NLSM. 
  In panel (a) the boundary solid (dashed) lines represent the merging of the interface states' band with the upper (lower) FWSM bulk eigenbands. The red (blue) lines describe the merging into the conduction (valence) band, and red and blue dots represent WPPs of positive and negative chirality, respectively. The Fermi line $\mathcal{E}=0$ is shown in black in panel (a) and in orange line in panel (b). The parameters used are: $u=1,\Omega=10$, $\Lambda_{\rm U}=2$, $\Lambda_{\rm L}=3$, $\phi_{\rm U} = \pi$, $ \phi_{\rm L}=0$.
  \label{fig:NonSymmetricNumeric}}
\end{figure}
\begin{figure}[t] 
\centering
\begin{overpic}[height=6.45cm]{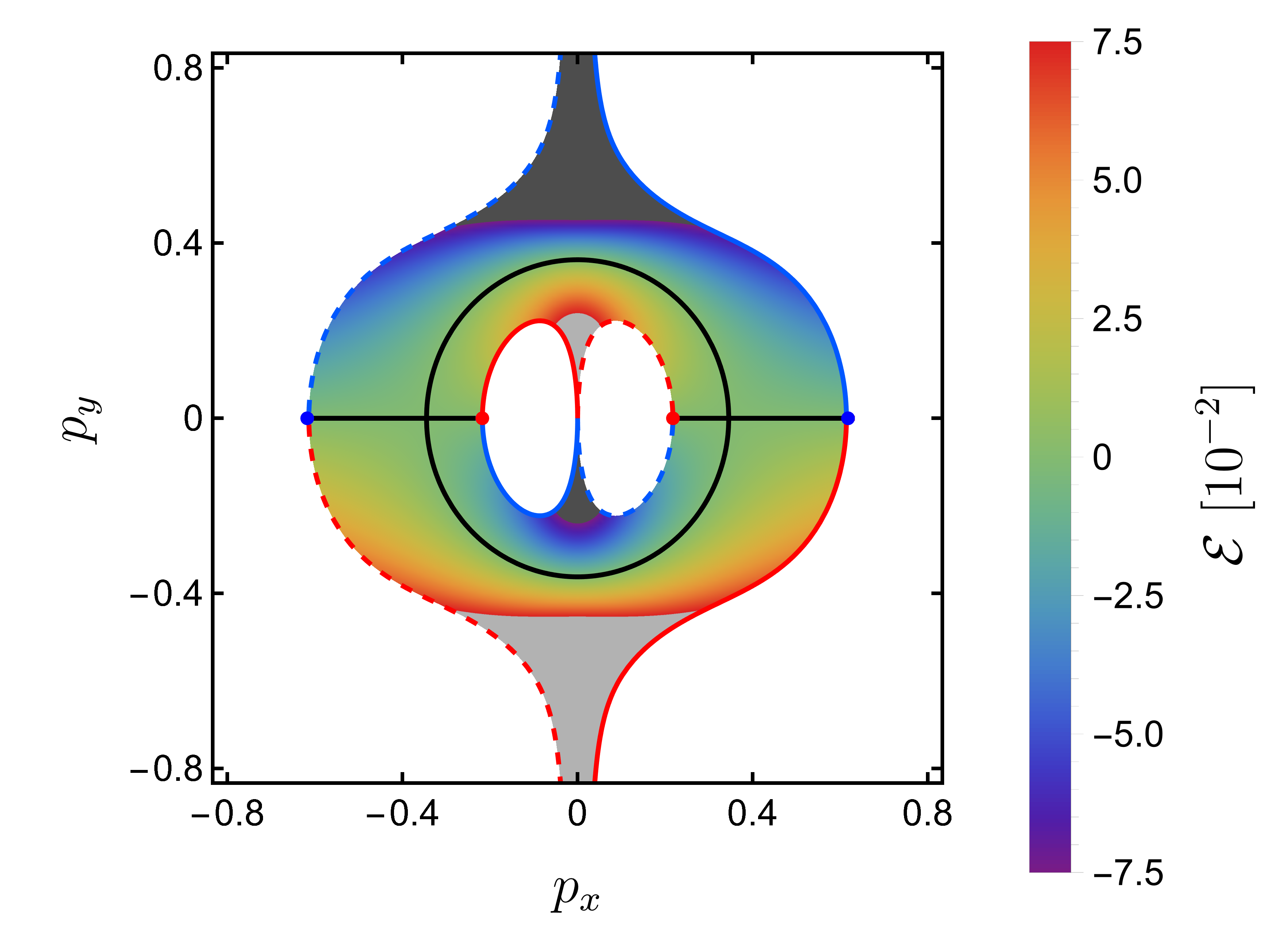}\put(4,75){(a)}\end{overpic}\\
\hspace{-6.30em}\begin{overpic}[height=6.15cm]{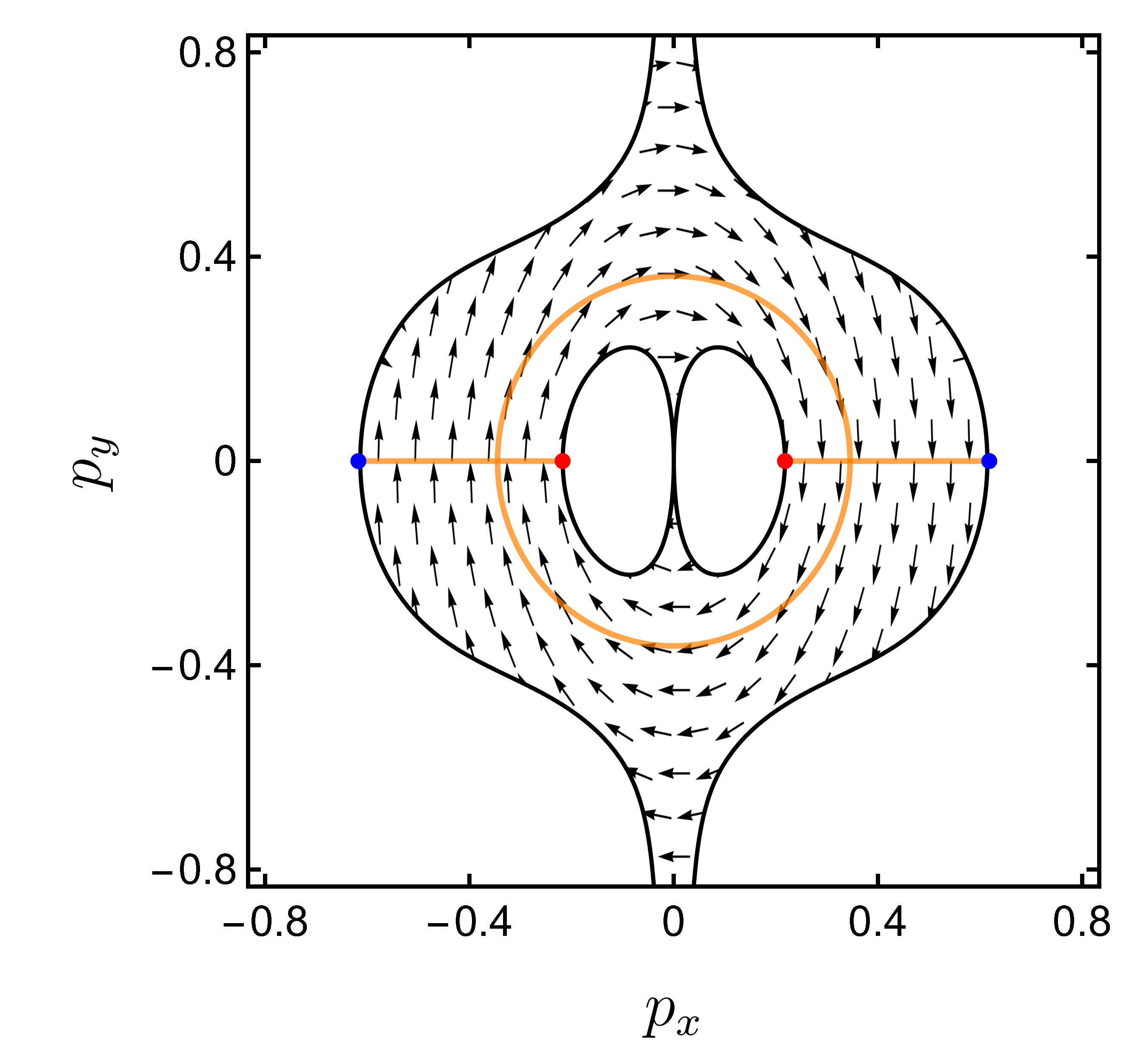}\put(4,95){(b)}\end{overpic}
  \caption{(a) Density plot of the energy dispersion, and (b) pseudo-spin texture plot of the interface states as a function of the momentum components $p_x$ and $p_y$,  with $\Lambda_{\rm U}=\Lambda_{\rm L}$ and in the presence of a delta-like magnetic field along the interface $z=0$, by holding the quadratic term in $p_z$ in the Hamiltonian~\eqref{eq:H_NLSM_Dimensionless_Numerical} which describes the NLSM. 
  In panel (a) the boundary solid (dashed) lines represent the merging of the interface states' band with the lower (upper) FWSM bulk eigenbands. The red (blue) lines describe the merging into the conduction (valence) band, and red and blue dots represent WPPs of positive and negative chirality, respectively. Positive (negative) out of range values are depicted as light (dark) gray areas, and the Fermi line $\mathcal{E}=0$ is shown in black in panel (a) and in orange line in panel (b). The parameters used are: $u=1,\Omega=10$, $\Lambda_{\rm U}=\Lambda_{\rm L}=3$, $\phi_{\rm U} = \pi$, $ \phi_{\rm L}=0$, and the magnetic parameter is set at $p_0 = 1/5$.
  \label{fig:MagneticNumeric}}
\end{figure}
In this appendix, we present the numerical results for the energy dispersion and the pseudo-spin texture of the interface system obtained from a two bands continuous model of the NLSM by including the quadratic momentum contribution along all directions. 
We obtain qualitative agreement with the findings of the main text, where we found analytical results by instead neglecting the quadratic term in $p_z$.
In particular, we show that the neglected term has no impact on the topological characterization of the interface system.
We describe the electronic system with the following Hamiltonian
\begin{equation}\label{eq:H_NLSM_Dimensionless_Numerical}
	\mathcal{H}\left(\bm{p}\right) = \left[1-\left(p_x^2 + p_y^2 + p_z^2\right)\right]\sigma_x + u p_z \sigma_z.
\end{equation}
Similarly to Sec.~\ref{sec:driving} of the main text,
we illuminate two regions of a NLSM with two beams of light, where each one is characterized by a vector potential of the form expressed in Eq.~\eqref{eq:VectorPotential}. 
Each side of the NLSM can be effectively described, in the high frequency regime, by the Floquet Hamiltonian~\cite{Yan_2016}
\begin{equation}\label{eq:H_FWSM_Numerical}
	 \mathcal{H}_{j}(\bm{p}) = \left[\bar{p}_j^2 - \left( p_x^2 +p_y^2 +p_z^2\right) \right] \sigma_x + \lambda_j p_y \sigma_y + u p_z \sigma_z,
\end{equation}
where
\begin{subequations}\label{eq:H_FWSM_Factors_LU_Numerical}
\begin{align}
    \bar{p}_j & = \sqrt{1- e^2 \Lambda_j^2} \label{eq:H_FWSM_Factors__LU_Numerical_a} ,\\
    \lambda_j & = - 2 e^2 \Lambda_j^2 u \cos{\left(\phi_j\right)} /\Omega ,\label{eq:H_FWSM_Factors__LU_Numerical_b}\\
     \Lambda_j&=\frac{A_j}{b},
\end{align}
\end{subequations}
with $j\in\{\rm L,U\}$ and $b=\sqrt{m/B}$. 
A quantitative difference with the model used in the main text
is visible in the definition of the term $\bar{p}_j$, in fact in Eq.~\eqref{eq:H_FWSM_Factors__LU_a} of the main text one has $\bar{p}_j= \sqrt{1- e^2 \Lambda_j^2/2}$. 
We follow the algebraic procedure used in Sec.~\ref{sec:interface}. Firstly, we solve the eigenvalue problem for each half-space by using the ansatz shown in Eq.~\eqref{eq:Ansatz}. Afterwards, we impose the continuity of the wavefunction and of its derivative at the interface $z=0$. Finally, we force the normalizability of the wavefunction in the entire space.
These conditions lead to a problem similar to that shown in Sec.~\ref{sec:interface}, 
but here the matrix ${\cal M}$ becomes four-dimensional.
The doubling of the dimension of the matrix ${\cal M}$ makes the problem hard for an analytical approach, then we face it with a numerical approach~\cite{Mathematica_123}.

In the symmetric case ($\Lambda_{\rm U} = \Lambda_{\rm L}$), the energy dispersion of the interface electron system has the same features of the analytical expression shown in Fig.~\ref{fig:EnergyDispersionNonMagnetic}~(a). Here, the interface states are defined in the whole 2D momentum space with the exception of the $p_y=0$ axis, where the interface band merges with the bulk bands, and the Fermi line at $\mathcal{E}=0$ is composed by two semicircumferences that meet at the WPPs. 
Moreover, by using the definition of the pseudo-spin $\left\langle \bm{\sigma} \right\rangle_{\bm{p}_\bot}$ in Eq.~\eqref{eq:Pseudospin_definition}, one finds that the pseudo-spin has a trivial pattern, in particular, it is parallel (antiparallel) to the $\hat{x}$ axis for $p_y>0$ ($p_y<0$).
 
We now consider the interface system obtained in the asymmetric case, namely the intensities of the light beam are different ($\Lambda_{\rm U} < \Lambda_{\rm L}$).
Figure~\ref{fig:NonSymmetricNumeric} shows the energy dispersion, panel (a), and the pseudo-spin texture, panel (b), of the interface states.
We have chosen the same values of the parameters $u,\Omega,\Lambda_j,\phi_j$ used in Sec~\ref{sec:interface_energy} of the main text.
In Fig.~\ref{fig:NonSymmetricNumeric}~(a) the interface states 
are defined in a region of the 2D momentum space delimited by the solid (dashed) lines which represent a merging of the interface band with the FWSM bulk bands delocalized along the upper region $z>0$ (lower region $z<0$). 
Moreover, red (blue) lines represent a merging of the interface band with a conduction (valence) band, while red (blue) dots are WPPs of positive (negative) chirality. 
The pseudo-spin texture of Fig.~\ref{fig:NonSymmetricNumeric}~(b) shows a trivial pattern in terms of the winding number defined in Eq.~\eqref{eq:WindingNumber_Definition}.
Apart from quantitative differences, mainly given by the modifications of the WPPs, both the energy dispersion and the pseudo-spin texture are qualitatively unaffected by the presence of the quadratic $p_z$ term in the Hamiltonian \eqref{eq:H_NLSM_Dimensionless_Numerical}.

To conclude, we introduce the delta-like magnetic barrier at the interface $z=0$ described by Eq.~\eqref{eq:MagneticBarrier}. 
Figure~\ref{fig:MagneticNumeric}~(a) is the density plot of the energy dispersion of the interface states, while Fig.~\ref{fig:MagneticNumeric}~(b) represents the pseudo-spin texture, the parameters are set at the same values used in Sec~\ref{sec:interface_magnetic}.
Here, we obtain that both the energy dispersion relation and the pseudo-spin texture are qualitatively in agreement with the analytical analysis of the main text.
In particular, the non-vanishing winding number of the pseudo-spin texture defined in Eq.~\eqref{eq:WindingNumber_Definition} occurs exactly like in the main text.

\section{Winding number and pseudo-spin texture}\label{sec:numerical}

In the main text we characterize the topology of the interface system through the winding number of Eq.~\eqref{eq:WindingNumber_Definition} as
\begin{equation}\label{eq_App:WindingNumber_Definition}
\nu\left(\Gamma\right) = \frac{1}{2\pi} \int_{0}^{1} \left( \left\langle \bm{\sigma} \right\rangle_{\bm{p}_\perp} \left( \tau \right) \times \frac{d}{d\tau} \left\langle \bm{\sigma} \right\rangle_{\bm{p}_\perp} \left( \tau \right) \right)_z d\tau,
\end{equation}
where $\tau \in \{ 0, 1 \} $ parametrizes a closed path $\Gamma$ in momentum space that surrounds the origin of momenta. Equation~\eqref{eq_App:WindingNumber_Definition} is intuitive in that it simply keeps track of whether the pseudo-spin is rotating clockwise or counterclockwise during its evolution in $\tau$. Mathematically, since the pseudo-spin is a normalized quantity we can write it as $\left\langle \bm{\sigma}\right\rangle_{\bm{p}_{\perp}}  \left(\tau\right) = \left(\cos \theta\left(\tau\right), \sin \theta\left(\tau\right) \right)$ (neglecting the irrelevant $z$ component), where $\theta\left(\tau\right)$ defines the direction of the pseudo-spin in the $x-y$ plane. Substituting this expression in Eq.~\eqref{eq_App:WindingNumber_Definition} we obtain
\begin{equation}\label{eq_App:WindingNumber_Calculations}
\begin{aligned}
	\nu\left(\Gamma\right) = & \frac{1}{2\pi} \int_{0}^{1} \frac{d \theta}{d \tau} \left(\cos \theta\left(\tau\right), \sin \theta\left(\tau\right) \right) \times \\ \times & \left(-\sin \theta\left(\tau\right), \cos \theta\left(\tau\right) \right) d\tau \nonumber \\ = & \frac{1}{2\pi} \left[ \theta\left(1\right) - \theta\left( 0 \right) \right] = \nu,
\end{aligned}
\end{equation}
where in the last line we have used the fact that $\Gamma$ is a closed curve and as such $\theta\left(1\right)-\theta\left(0\right)= 2\pi \nu$.

The pseudo-spin vector is defined through Eq.~\eqref{eq:Pseudospin_definition}:
%
\begin{align}\label{eq_App:Pseudospin_Simplified}
	\left\langle \bm{\sigma}\right\rangle_{\bm{p}_\perp} & = \int d\bm{r} \Psi_{\bm{p}_\perp}^\dagger(\bm{r}) \bm{\sigma} \Psi_{\bm{p}_\perp}(\bm{r}) \nonumber \\
	& = \mathcal{N}_0 \begin{pmatrix}	\psi_{1,-}^{U} & \psi_{2,-}^{U}	\end{pmatrix}^* \bm{\sigma} \begin{pmatrix}	\psi_{1,-}^{U} \\ \psi_{2,-}^{U}	\end{pmatrix} \nonumber \\ & = 2 \mathcal{N}_0 \left( \rm{Re}\left(\psi_{1,-}^{U*} \psi_{2,-}^{U}\right),\rm{Im}\left(\psi_{1,-}^{U*} \psi_{2,-}^{U}\right)\right),
\end{align}
where $\mathcal{N}_0 = \begin{pmatrix}	\psi_{1,-}^{U} & \psi_{2,-}^{U}	\end{pmatrix}^{*} \begin{pmatrix}	\psi_{1,-}^{U} & \psi_{2,-}^{U}	\end{pmatrix}^T $, $\bm{\sigma} = \left( \sigma_1, \sigma_2 \right)$ (as said before, we neglect the $z$ component), and where we could use only the upper space spinor because of the continuity condition at the interface.
For the asymmetric case of Sec.~\ref{sec:interface_pseudospin}, substituting Eq.~\eqref{eq:Spinors} into Eq.~\eqref{eq_App:Pseudospin_Simplified} and calculating the result along the circumference at zero energy shown in Fig.~\ref{fig:PseudoSpinNonSymmetric} and parametrized by $(p_x,p_y)= R (\cos(\theta),\sin(\theta))$, where $R=\sqrt{\left(\lambda_U \bar{p}^2_L - \lambda_L \bar{p}^2_U\right) / \left(\lambda_U -\lambda_L\right)}$, we get
\begin{equation}\label{eq_App:Pseudspin_Simplified_Asymmetric}
	\left\langle \bm{\sigma} \right\rangle_{\bm{p}_{\perp}} = \frac{1}{\sqrt{\left( \lambda_U R \sin \theta\right)^2 + \left(R^2 - \bar{p}_U^2\right)^2}} \left(\lambda_U R \sin \theta, R^2 - \bar{p}_U^2\right),
\end{equation}
which is the normalized version of Eq.~\eqref{eq:Circular_Arc_Pseudospin}; $\lambda_j$ and $\bar{p}_j$ are defined in Eqs.~\eqref{eq:H_FWSM_Factors_LU}. Finally, inserting Eq.~\eqref{eq_App:Pseudspin_Simplified_Asymmetric} into Eq.~\eqref{eq_App:WindingNumber_Definition}, the resulting winding number is vanishing
\begin{equation}\label{eq_App:WindingNumber_NonSymmetric}
	\nu = - \frac{R^2-\bar{p}_U^2}{2\pi} \int_{0}^{2\pi} \frac{\lambda_U R \cos \theta}{\left(\lambda_U R \sin \theta\right)^2 + \left(R^2 - \bar{p}_U^2\right)^2}d\theta=0.
\end{equation}

On the contrary, performing the same calculations for the magnetic case of Sec.~\ref{sec:interface_magnetic} along the circumference at zero energy shown in Fig~\ref{fig:Magnetic}~(b), which is parametrized by $(p_x,p_y)=\bar{R}(\cos(\theta),\sin(\theta))$, where $\bar{R} = \sqrt{\bar{p}^2-p_0^2}$, we obtain
\begin{equation}\label{eq_App:Pseudspin_Simplified_Magnetic}
	\left\langle \bm{\sigma}\right\rangle_{\bm{p}_{\perp}} = \frac{1}{\left( \lambda_U \sin \theta\right)^2 + \left(2 p_o \cos \theta\right)^2} \left(\lambda_U \sin \theta, -2 p_0 \cos \theta\right)
\end{equation}
for the pseudo-spin vector, and
\begin{equation}\label{eq_App:WindingNumber_Magnetic}
	\nu = \frac{\lambda_U p_0}{\pi} \int_{0}^{2\pi} \frac{1}{\left(\lambda_U \sin \theta\right)^2 + \left( 2 p_0 \cos \theta \right)^2} d \theta = +1
\end{equation}
for the winding number.

\section{Non-triviality condition}\label{sec:NonTrivialityCondition}

In Sec.~\ref{sec:interface_magnetic} of the main text, we mainly focused on the setup with a magnetic barrier between two FWSMs which are illuminated by light beams with opposite polarizations and same intensities. If we also allow for the intensities to be different on the two FWSMs then the energy dispersion of the interface system will be expressed as
\begin{widetext}
\begin{equation}\label{eq_App:Energy_MagneticAsymmetric}
	\mathcal{E}\left(\bm{p}_{\perp}\right) = p_y \frac{\lambda_U \bar{p}_L^2 - \lambda_L \bar{p}_U^2 - \left[ \left( p_x^2 + p_y^2 + p_0^2\right)\left(\lambda_U-\lambda_L\right)+2 p_0 p_x \left(\lambda_U + \lambda_L\right)\right]}{\sqrt{\left( \bar{p}_U^2 - \bar{p}_L^2 + 4 p_0 p_x\right)^2 + \left[p_y\left(\lambda_U-\lambda_L\right)\right]^2}},
\end{equation}
\end{widetext}
which has to be compared with Eqs.~\eqref{eq:Solution_Energy_Light} and \eqref{eq:EnergyDispersion_Magnetic_Simplified}. Similarly to the main text, see Fig.~\ref{fig:Magnetic}, the circumference at zero energy, $\mathcal{E}\left(\bm{p}_{\perp}\right)=0$, is now given by:
\begin{equation}\label{eq_App:MagneticNonSymmetric_CircularArc}
\begin{aligned}
	p_x & = p_c + \tilde{R} \cos \theta, \\
	p_y & = \tilde{R} \sin \theta,
\end{aligned}
\end{equation}
with
\begin{subequations}\label{eq_App:NonSymmetricMagneticCircumference}
\begin{align}
	p_c & = -p_0 \frac{\lambda_U + \lambda_L}{\lambda_U-\lambda_L}, \label{eq_App:NonSymmetricMagneticCircumference_C} \\
	\tilde{R}^2 & = \frac{\lambda_U \bar{p}_L^2 - \lambda_L \bar{p}_U^2}{\lambda_U-\lambda_L} + \frac{4\lambda_U \lambda_L}{\left(\lambda_U -\lambda_L\right)^2} p_0^2. \label{eq_App:NonSymmetricMagneticCircumference_R}
\end{align}
\end{subequations}
Performing the same calculations as in Appendix~\ref{sec:numerical}, the pseudo-spin vector along this curcumference can be expressed as
\begin{equation}\label{eq_App:Pseudspin_Simplified_MagneticNonSymmetric}
\begin{aligned}
	\left\langle \bm{\sigma}\right\rangle_{\bm{p}_{\perp}} & = \frac{1}{\sqrt{\left(\lambda_U \tilde{R}\sin\theta\right)^2 + \left(\tilde{R}^2 + p_c^2 -\bar{p}_U^2 - 2 R p_c \cos \theta\right)^2}} \times \\ & \times \left(\lambda_U \tilde{R} \sin \theta, \tilde{R}^2 + p_c^2 -\bar{p}_U^2 - 2 \tilde{R} p_c \cos \theta \right).
\end{aligned}
\end{equation}
Hence, the non-triviality condition of the pseudo-spin pattern is given by the "rotation" of the second component of the pseudo-spin vector in Eq.~\eqref{eq_App:Pseudspin_Simplified_MagneticNonSymmetric}, namely:
\begin{equation}\label{eq_App:NonTrivialityConditionPre}
	\left| \tilde{R}^2 + p_c^2 -\bar{p}_U^2 \right| < \left| 2 \tilde{R} p_c \right|.
\end{equation}
It can then be numerically checked that Eq.~\eqref{eq_App:NonTrivialityConditionPre} is satisfied when:
\begin{equation}\label{eq_App:NonTrivialityCondition}
	\left| \bar{p}_{\rm U} - \bar{p}_{\rm L} \right| < 2\left| p_0 \right| < \bar{p}_{\rm U} + \bar{p}_{\rm L}~,
\end{equation}
which is Eq.~\eqref{eq:NonTrivialityCondition}.

\end{document}